\documentclass[10pt]{iopart}

\usepackage{graphicx}
\usepackage{amssymb}
\usepackage{float}

\newtheorem{theorem}{Theorem}[section]
\newtheorem{defin}[theorem]{Definition}
\newtheorem{lemma}[theorem]{Lemma}
\newtheorem{prop}[theorem]{Proposition}

\newcommand{\del}{\partial}
\newcommand{\eps}{\varepsilon}

\newcommand{\dd}[2]{\frac{\del #1}{\del #2}}
\newcommand{\ddeval}[3]{\left.\dd{#1}{#2}\right|_{#3}}

\newcommand{\DD}[2]{\frac{\rmd #1}{\rmd #2}}
\newcommand{\DDeval}[3]{\left.\DD{#1}{#2}\right|_{#3}}

\newenvironment{algo}{%
  \begin{list}{}{%
      \setlength{\topsep}{0cm}
      \setlength{\partopsep}{0cm}
      \setlength{\itemsep}{0cm}
      \setlength{\parsep}{0cm}
      \setlength{\leftmargin}{1cm}}
    }{
  \end{list}
  }

\begin{document}
\title[Computation of CMC surfaces]{Numerical computation of constant
mean curvature surfaces using finite elements} 
\author{Jan Metzger}
\address{Max-Planck-Institut f{\"u}r Gravitationsphysik,
Albert-Einstein-Institut, Am M{\"u}hlenberg 1, D-14476 Golm,
Germany\footnote{The manuscript of this paper was prepared while the
author was financed by the Sonderforschungsbereich 382 at the
Universit{\"a}t T{\"u}bingen. This mansuscript has the preprint number
AEI-2004-062.}}
\ead{jan@everest.mathematik.uni-tuebingen.de}
\begin{abstract}
  This paper presents a method for computing two-dimensional constant mean
  curvature surfaces. The method in question uses the variational aspect of
  the problem to implement an efficient algorithm. In principle it is a flow
  like method in that it is linked to the gradient flow for the area
  functional, which gives reliable convergence properties. In the background a
  preconditioned conjugate gradient method works, that gives the speed of a
  direct elliptic multigrid method.
\end{abstract}
\submitto{\CQG}
\maketitle
%
\section{Introduction}
\label{s:intro}
The computation of surfaces with prescribed mean curvature is an important
problem in numerical relativity with an abundance of applications. We
will only give a few examples here, which we had in mind while
developping this algorithm. 

The first example are marginally trapped surfaces, that is closed
spherical surfaces $\Sigma$ in a Riemannian 3-manifold $(M,g)$, such
that $H \pm P = 0$ on $\Sigma$. Here $H$ denotes the mean curvature of
$\Sigma$ and $P= \textrm{tr}_\Sigma K$ is the trace of $K$ along
$\Sigma$. The extra tensor field $K$ on $M$ represents the second
fundamental form of $M$ in spacetime. Apparent horizons are outermost
marginally trapped surfaces and are used in numerical simulations for
various purposes like black hole location or excision of black holes
in numerical simulations inside their apparent horizon. For an
introduction and furter references see for example \cite{DKSS:2004,BS:2003}.

In the special, time symmetric case $K=0$, i.e. $P=0$, these surfaces are
minimal surfaces, that is critical points of the area functional.  When
surfaces of minimal area are considered that satisfy the additional
constraint, that they include a given volume with the minimal surface, the
minimal surface can be ``blown up'' to render an evenly spaced, geometrically
defined foliation of the exterior domain, as it is explained in Huisken and
Yau \cite{HY:1996} and Huisken \cite{H:1998}. The surfaces of this foliation
have constant mean curvature
\[ H = \textrm{const}\,. \]
This the Euler-Lagrange equation to the constrained problem. It is a
quasilinear, second order, degenerate elliptic equation for the position of
the surface. The size of the enclosed volume, as well as the area of the
constant mean curvature surfaces is then a generic candidate for a geometric
radial coordinate.

Besides this, there are more applications for constant mean curvature
foliations.  One of them is to define a concept for the center of
mass of an isolated gravitating system \cite{HY:1996}. Another application is
that a geometrically defined foliation also can be used to construct
geometric gauge conditions for time evolution in the Einstein equations as
described in \cite{KH:1995}. In addition the proof of the Riemannian Penrose
inequality by Bray \cite{Bray:1997} uses isoperimetric surfaces, which are
special cases of constant mean curvature surfaces. This proof establishes the
monotonicity of the Hawking mass on these surfaces with increasing enclosed
volume, given a positive energy condition. This energy condition translates
into the geometric condition for $M$ to have nonnegative scalar curvature. The
numerically computed constant mean curvature surfaces may therfore be used to
detect and measure gravitational fields.

A considerable number of methods have been exploited to find apparent
horizons, a current overview of which is given in Thornburg
\cite{Thornburg:1996}. Baumgarte and Shapiro \cite{BS:2003} also
review some techniques for locating apparent horizons. Schnetter
\cite{Schnetter:2003} computed the more general ``constant expansion
surfaces'' with $H\pm P = \textrm{const}$. These methods can be
classified into three different approaches, namely flow like methods
using a curvature flow to locate the apparent horizon, direct methods
using a Newton method to solve the elliptic apparent horizon equation,
and indirect minimization methods that try to minimize an $L^2$-error
integral.

The flow like methods are very robust as they converge for a large set
of initial data and are able to model the change of topology by using
level set methods \cite{Tod:1991,SHM:2000}. However, these methods are
rather slow. The more and more popular direct elliptic methods are
very fast but have a small domain of convergence as is noted by
Thornburg \cite{Thornburg:1996}. The minimization methods are
problematic, since minimizing the $L^2$-error functional corresponds
to solving a fourth order PDE, while the actual problem is only a
second order PDE\@. This does not only spoil the condition of the
problem, but may also introduce invalid solutions.  All these methods,
however, share the fact that they use finite differencing.

In contrary, this article describes a \emph{finite element} based \emph{direct}
minimization method to compute constant mean curvature surfaces
\[ H = \textrm{const}\,, \]
that inherits it the big domain of convergence from the flow like method,
while rivaling the speed of the direct elliptic method. The method as it is
presented here can not be applied without modification to horizon finding in
the general non-time symmetric case with $K\not\equiv 0$, but an appropriate
modification is outlined in section~\ref{s:generalized}.

We consider $\mathbf{R}^3$ to be equipped with a general metric $g_{ij}$. For
the purposes of general relativity this metric will be asymptotically flat,
that is in rectangular coordinates the metric components satisfy 
\[ g_{ij} = \delta_{ij} + \mathcal{O}(r^{-1})\,. \]
We also will consider two-dimensional spherical surfaces $\Sigma\subset\mathbf{R}^3$.
The induced metric on $\Sigma$ will be denoted by $\gamma$, the outer normal
by $\nu$ and the second fundamental form $A =\nabla\nu$. The mean curvature is
labeled $H=\textrm{tr}_\Sigma A$. In the following we use Einstein's summation
convention such that Latin indices range from 1 to 3 whereas Greek indices
range from 1 to 2. We will frequently use the abbreviation CMC for ``constant
mean curvature''.

Section~\ref{s:background} will explain the relationship of constant mean
curvature surfaces to the isoperimetric problem and give some theoretical
results concerning existence and properties of constant mean curvature
surfaces. The algorithm is explained in section~\ref{s:algo} and some numerical
examples are given in section~\ref{s:examples}.

%
\section{Theoretical Background}
\label{s:background}
This paper uses the fact that constant mean curvature surfaces are critical points
of the isoperimetric problem. The isoperimetric problem is to find a surface
enclosing a given volume that has minimal area. Formulated precisely, this is
\begin{defin} Let $M$ be a Riemannian manifold, then a compact subset
  $\Omega\subset M$ is called a solution to the isoperimetric problem if for
  all $\Omega'\subset M$ with $\textrm{Vol}\,(\Omega') = \textrm{Vol}\,(\Omega)$ the inequality
\[ |\del \Omega| \leq |\del\Omega'| \]
holds.
\end{defin}
To treat this problem with tools from the calculus of variations one denotes by
a variation in $M$ a smooth map $F:M\times(-\eps,\eps)\rightarrow M$
such that $F(\cdot,0):M\rightarrow M$ is the identity and for all $t\in
(-\eps,\eps)$ the map $F(\cdot,t):M\rightarrow M$ is a diffeomorphism.
We collect the following facts from the literature, see eg. \cite{BC:1984} for
a flat background metric or \cite{BC:1988} in the general case.
\begin{lemma}
  \label{lemma:gradients}
  Given any set $\Omega\subset M$ with smooth boundary $\Sigma=\del\Omega$ and a
  variation $F$ with normal velocity $f=g\left(\ddeval{F}{t}{t=0},\nu\right)$
  on $\Sigma$, the following variation formulas for the volume $\textrm{Vol}\,(\cdot)$ of $\Omega$ and
  area $|\cdot|$ of $\Sigma$ hold.
  \begin{eqnarray*}
    \DDeval{}{t}{t=0}\textrm{Vol}\,(F(\Omega,t)) &=& \int_\Sigma f\,\rmd\mu_\Sigma \\
    \DDeval{}{t}{t=0} |F(\Sigma,t)| &=& \int_\Sigma Hf\,\rmd\mu_\Sigma 
  \end{eqnarray*}
\end{lemma}
Therefore $H$ can be interpreted as $L^2$-gradient of the area functional, and the
constant function $1$ as the $L^2$-gradient of the volume functional. We have
the following characterization of volume preserving variations.
\begin{lemma}
  \label{lemma:volpres}
  If $F$ is a variation that preserves the volume of $\Omega\subset M$, then
  $\int_\Sigma fd\mu_\Sigma=0$ for $\Sigma=\del\Omega$ and $f$ the normal
  velocity of $F$ on $\Sigma$.  
  Conversely, if  $f$ is a function with $\int_\Sigma fd\mu_\Sigma=0$ then there
  exists a volume preserving variation with $f$ as normal velocity.
\end{lemma}
Introducing the usual Lagrangian, the Euler-Lagrange equation of the
isoperimetric problem can be computed.
\begin{prop}
  If $\Omega$ is a smooth solution to the isoperimetric problem, then $\Omega$ is
  bounded by a constant mean curvature surface.
\end{prop}
This only characterizes critical points of the isoperimetric problem. To give
a better description the usual concept of \emph{stability} has to be
introduced.
\begin{defin}
  A constant mean curvature surface is called \emph{stable} if the second
  variation of area in volume preserving directions is nonnegative, and
  \emph{strictly stable} if it is positive.
\end{defin}
Due to the characterization of volume preserving variations in
lemma~\ref{lemma:volpres}, a sufficient condition for stability is the
nonnegativity of the Jacobi operator
\[ Jf = -\Delta f - f\left(|A|^2+\textrm{Ric}(\nu,\nu) \right) \]
that is the inequality
\[ \int_\Sigma f^2\left(|A|^2+\textrm{Ric}(\nu,\nu)\right)\,\rmd\mu  
\leq \int_\Sigma |df|^2\,\rmd\mu \] 
for all $f\in C^\infty(\Sigma)$ with $\int
f=0$.  A constant mean curvature surface $\Sigma$ is strictly stable, if there exist
$\alpha>0$ such that
\[ \alpha\int_\Sigma f^2\,\rmd\mu \leq \int_\Sigma fJf\,\rmd\mu \] 
for all $f$ with $\int f=0$. A strictly stable constant mean curvature surface is an
isolated local minimum of the isoperimetric problem, in the sense that there
is no volume preserving variation that does not increase area.

The algorithm presented was constructed having in mind spatial slices of
isolated gravitating systems in general relativity. These slices have (in
absence of linear momentum) the following asymptotic behavior of the metric.
\begin{defin} 
  \label{def:asympt_flat}
  A \emph{strongly asymptotically flat manifold} is a
  Riemannian manifold $M$ together with a metric $g$, such that there is a compact
  set $C\subset M$ and a diffeomorphism 
  $x: M\setminus C \rightarrow \mathbf{R}^3\setminus B_R(0)$ for some $R$ and, such
  that, in the coordinates given by $x$, the metric $g$ has the form
  \[ g_{ij}=\left(1+\frac{m}{2r}\right)^4\delta_{ij} +
  Q_{ij} \]
  with the following decay conditions
  \[ |Q_{ij}| \leq Cr^{-2} \quad |\partial^l Q_{ij}|\leq
  Cr^{-2-l}\quad l=1,2,3,4\]
\end{defin}
In this setting Huisken and Yau \cite{HY:1996} have proved the following
\begin{theorem}
  \label{thm:hy}
  Let $(M,g)$ be a strongly asymptotically flat manifold with $m>0$. Then
  there exists a compact set $C\subset M$ such that on $M\setminus C$ exists a unique
  foliation by spherical constant mean curvature surfaces, such that
  \begin{enumerate}
  \item for growing radius these surfaces approximate Euclidean spheres, 
  \item the centers of these spheres converge to a point in $\mathbf{R}^3$,
  \item and the respective surfaces are strictly stable with respect to the
    isoperimetric problem.
  \end{enumerate}
\end{theorem}
For the purposes of this article, the way Huisken and Yau establish the
existence of constant mean curvature surfaces is very interesting. They use the
\emph{volume preserving mean curvature flow}. Solving this flow means finding a map
$F:\Sigma_0\times(0,T)\rightarrow N$ for an initial surface $\Sigma_0\subset
M$ with
\begin{eqnarray*}
  \dd{F}{t} &=& (h-H)\nu \qquad \textrm{for}\qquad t \geq 0\\
  F(0) &=& \Sigma_0 
\end{eqnarray*}
where $h = |\Sigma_t|^{-1} \int_{\Sigma_t} H d\mu_t$ with $\Sigma_t=
F(\Sigma_0,t)$. Huisken and Yau show that in the above setting a solution
exists for all times, in case the flow is started from a Euclidean sphere of
radius bigger than some critical radius. For $t\rightarrow\infty$ the surfaces
$\Sigma_t$ then converge to a surface with constant mean curvature.

In view of the variational perspective, this is the flow to the gradient of
the area functional projected onto the volume preserving variations, which is
a technique for solving constrained minimization problems. Huisken and Yau
show that this method converges. The key issue in this proof is the stability
of the CMC-surfaces. That is, these surfaces are local minimizers of the
isoperimetric problem.

These facts enable us to use a constrained minimization algorithm to approach
the numerical computation of constant mean curvature surfaces, since the most
efficient of these methods rely on the positivity of the second variation. 
%
\section{Description of the Algorithm}
\label{s:algo}
As explained before, Huisken and Yau \cite{HY:1996} have shown that a
projected gradient flow method for constrained minimization converges in the
analytic case. From the numerical viewpoint there are much better methods for
solving such problems, since projected gradient methods tend to converge
slowly and do not preserve the constraint during iteration.

The approach taken here is to convert the constrained problem into an
unconstrained minimization problem, which is then solved by a preconditioned
conjugate gradient method.

\subsection{The surface model}
\label{s:surfacemodel}
Discrete surfaces are modeled as a triangulated meshes, that is sets of
vertices, edges and triangles linked together according to the topology of the
triangulation. 

To each vertex $p$ of a given triangulation $\tilde\mathcal{T}$ a variation vector
$r(p)$ is attached. A vertex can only move into this direction. Associated to
$\tilde\mathcal{T}$ is a discrete set of linear finite elements
\[ \mathcal{S} = \left\{ \sum_{p\in\mathcal{T}} \alpha_p \phi_p : \alpha_p\in\mathbf{R} \right\} \]
where $\phi_p(q) = \delta_{pq}$ for all vertices $q$ of the triangulation and
$\phi_p$ restricted to a triangle is linear.  As indicated in the definition,
the functions $\phi_p$ form a basis of $\mathcal{S}$, called the \emph{nodal
  basis}. A function $u\in\mathcal{S}$ is therefore characterized by its
values in the vertices of the triangulation.

Given any $u\in\mathcal{S}$ the graph of $u$ over $\tilde\mathcal{T}$ is the
triangulation $\mathcal{T}$ with the vertices $p=\tilde p+u(\tilde p)r(\tilde
p)$. Fix a reference triangulation $\tilde\mathcal{T}$ and model the unknown
surface as graph of a finite element function over $\tilde\mathcal{T}$.

For the purposes of applying hierarchical basis preconditioning, the
implementation uses a hierarchical data structure described by Leinen
\cite{Leinen:1995}. This object oriented approach stores the triangles in a
tree, the roots being coarse triangles with their subtriangles as child nodes
whenever the coarse triangle is divided. The edges form a similar structure.
To link edges and triangles, these two trees have cross references according
to their topology, such as triangles consisting of edges and edges being part
of triangles.  

An excellent outline of the method of finite elements and issues of
numerical integration as needed in the next section is given in
\cite{CL:1991}.

\subsection{Discrete Area and Volume Functionals}
To compute the area of a triangulation it is clearly possible to compute the
area of each triangle and sum up. The computation of the area of a single
triangle takes place in a standard situation. Define the \emph{standard
  triangle} $T_S\subset \mathbf{R}^2$ as the set
\[ T_S=\left\{(x_1,x_2)\in\mathbf{R}^2:x_1,x_2\geq 0, x_1+x_2\leq1\right\} \]
Every triangle $T=\Delta(p_0,p_1,p_2)\subset\mathbf{R}^3$ of the triangulation is diffeomorphic
to $T_S$ via the linear map
\[ F_T : T_S \rightarrow T : (x_1,x_2)\mapsto (1-x_1-x_2) p_0 + x_1 p_1 + x_2
p_2\,.\]
Then the area of $T$ is given by
\begin{equation}
  A(T) = \int_{T_S} \sqrt{\det(\gamma_{\alpha\beta})}\rmd x 
  \label{eq:area} 
\end{equation}
where $\gamma_{\alpha\beta}$ are the components of the induced metric on $T$
in the coordinates induced by $F_T$. If $\mathbf{R}^3$ was equipped with the
Euclidean metric, $\sqrt{\det(\gamma_{\alpha\beta})}$ would be a constant on
$T$. Then a quadrature formula to integrate (\ref{eq:area}) exactly is given
by
\[ A(T) = \frac{1}{2}\sqrt{\det\left(\gamma_{\alpha\beta}(1/3,1/3)\right)} \]
and this is the integration rule used to define the discrete area
\begin{equation}
  \hat A(T) = \frac{1}{2}\sqrt{\det\left(\gamma_{\alpha\beta}(1/3,1/3)\right)}\,.
  \label{eq:discrete-area}
\end{equation}
For this integration rule we obtain that 
\[ |A(T) - \hat A(T)| \leq c_0 h^2 A(T) \,,\]
where $h$ is the (Euclidean) diameter of $T$ and $c_0=c_0(g,\del g, \del^2 g)$
depends on the metric $g$, in particular, we have the bounds $c_0 \leq Cr^3$
in view of definition~\ref{def:asympt_flat}.
The above formula gives that for the whole triangulation we have
\[ |A(\mathcal{T}) - \hat A(\mathcal{T})| \leq c_0 h_{\textrm{max}}^2 A(\mathcal{T}) \]
where $h_{\textrm{max}}$ is the maximal diameter of the triangles of $\mathcal{T}$.
 
The discrete area $\hat A$ is as differentiable with respect to the values of
$u_p$ as the background metric $g$. Explicit formulas for the derivatives of
$\hat A$ can be computed. These formulas contain first derivatives of $g$ in
the $r(p)$ directions.

Consider for example the triangle $T_0$ with vertices $(p_0,p_1,p_2)$ and move
$p_1$ into the $r(p_1)=r_1$-direction, which gives a family of triangles
$T_\eps = \Delta(p_0, p_1+\eps r_1, p_2)$.  Then look at the following map
\[ F_\eps : T_S \rightarrow T_\eps : (x_1,x_2) \rightarrow
(1-x_1-x_2) p_0 + x_1 (p_1+ \eps r(p_1)) + x_2 p_2 \] 
and compute the $\eps$-derivative of the discrete area
expression in (\ref{eq:discrete-area}) using these coordinates. This gives 
\begin{eqnarray*}
  \DDeval{}{\eps}{\eps=0} \hat A(T_\eps) &=& \frac{1}{4\hat A(T_0)}
  \left(\gamma_{22}\,g(X_1,r_1) - \gamma_{12}\,g(X_2,r_1)\right) \\
  && + \frac{1}{24\hat A(T_0)}\left(\gamma_{11}\Lambda_{22,1} +
  \gamma_{22}\Lambda_{11,1} + 2\gamma_{12}\Lambda_{12,1}\right)\,.
\end{eqnarray*}
Here $X_\alpha = p_\alpha - p_0$, $\gamma_{\alpha\beta} = \gamma_{\alpha\beta}(1/3,1/3)$ and
\[ \Lambda_{\alpha\beta,\delta} = \ddeval{g_{ij}}{y^k}{F_0(1/3,1/3)}r^k_\delta X^i_\alpha
X^j_\beta\,. \]
Similar formulas can be derived for all vertices.

To define the discrete volume functional, a fixed reference triangulation $\tilde\mathcal{T}$
and a function $u$ is required. The discrete volume functional is defined as
the oriented discrete volume enclosed by the shell between $\tilde\mathcal{T}$ and the graph
of $u$ over $\tilde\mathcal{T}$. Orientation is chosen such that positive $u$ gives positive
volume and negative $u$ negative volume, that is the $r(p)$ are interpreted to
point outward.

\begin{figure}
  \centering
  \includegraphics[height=.6\linewidth,angle=270]{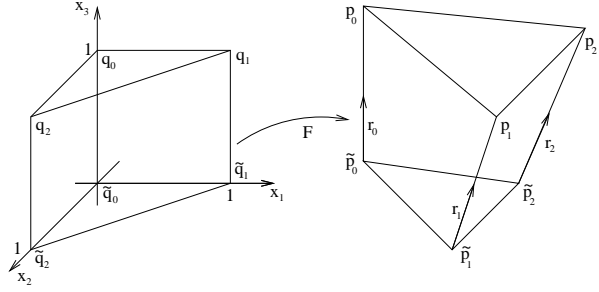}
  \label{fig:volcomp}
  \caption{Standard prism mapped to volume corresponding to one triangle.}
\end{figure}

For the computation of the volume corresponding to a single triangle, define
the standard prism $P_S = T_S\times[0,1]$. The prism $P_S$ can be mapped to
the volume $P$ between a triangle $\tilde T$ of $\tilde\mathcal{T}$ with vertices
$\tilde p_0,\tilde p_1,\tilde p_2$ and the corresponding triangle $T$ of
$\textrm{graph} u$ with vertices $p_i=\tilde p_i + u(\tilde p_i) r(\tilde p_i)$, $i=0,1,2$ via 
\begin{eqnarray*}
  G_P : P_S &\rightarrow& P \\
  (x_1,x_2,x_3) &\mapsto& \left(1-x_3\right) \left( \left(1-x_1-x_2\right)
    \tilde p_0 +  x_1 \tilde p_1 + x2 \tilde p_2 \right) \\
  &&+ x_3\left( \left(1-x_1-x_2\right) p_0 + x_1 p_1 + x_2 p_2 \right)
\end{eqnarray*}
as illustrated in figure~\ref{fig:volcomp} .
The oriented volume of $P$ is then given by
\begin{eqnarray*}
\int_P 1\,\textrm{dvol} &=& \int_P \sqrt{\det(g_{ij})}\,\rmd x \\
&=& \int_{P_S} \sqrt{\det(g_{ij}\circ G_P)} \det(d_x G_P)\,\rmd x\,.
\end{eqnarray*}
We choose the sign of this expression to match the orientation condition given
above. The integral in this formula is again replaced by a quadrature formula
which is exact in the case of a Euclidean background metric. Such a formula
can be constructed as product of a quadrature formula for $T_S$ and one for
$[0,1]$.  For $T_S$ we can use the center of gravity rule previously used for
the area. For $[0,1]$ an integration rule which is exact for polynomials of
degree two is sufficient. Such a rule is for example given by the two-point
Gauss rule with weights $1/2$ and evaluation points $1/2\pm 1/\sqrt{12}$ which
is exact for polynomials of degree three. The formula for discrete volume
computed with a general integration rule for $[0,1]$ with the $N$ weights
$a_k$ and evaluation points $z_k$ then reads
\begin{equation}
  \label{eq:discrete_volume}
\fl \hat V(P) = \sum_{k=1}^N  a_k
  \sqrt{\det(g_{ij}\left(G_P(1/3,1/3,z_k)\right)} 
  \det(\rmd_x G_P)(1/3,1/3,z_k) 
\end{equation}
There is no problem to evaluate these terms, but more elegantly $\det(\rmd_x G_P)(1/3,1/3,x_3)$
can be written as a polynomial in $u_p$ of degree up to three.  The six non-zero
coefficients of this polynomial can be computed from $\tilde\mathcal{T}$ only.
Therefore evaluation of this term only requires the evaluation of this
polynomial.

For this functional the error bounds 
\[ |V(T) - \hat V(T)| \leq c_1 h^2 U |V(T)|\, \]
hold, with analogous bounds for the volume of the whole triangulation. Here
$h$ is the maximum of the diameters of $T$ and $\tilde T$, $U = \max\{|u(p_i)
r(p_i)|:i=0,1,2\}$ and $c_1 = c_1(g,\del g, \del^2 g)$ depends on the metric
with $c_1 \leq Cr^{-3}$ in view of definition~\ref{def:asympt_flat}.

The discrete volume functional is again as differentiable with respect to the
nodal values of $u$ as the metric $g$. Explicit formulas for these derivatives
can be computed from (\ref{eq:discrete_volume}) in the same way as for
the area functional.

The discrete volume and area functionals are therefore constructed to give the
right answer in the Euclidean case since the asymptotically flat situation
suggests that this gives a good guess. Indeed, the error bounds given above
improve rapidly for growing radius. Their values and derivatives can be
computed per triangle using a fixed procedure, that is fixed time. Thus
evaluation of these two functionals takes time proportional to the number of
triangles.

Evaluation of the functionals and their derivatives requires the evaluation of
the metric and its derivatives at the requested points. Therefore, if the
metric is given on a grid it is necessary to interpolate these values.

Note that the gradient of the area functional is a discrete, weak analogon to
the mean curvature. This can be seen from lemma~\ref{lemma:gradients} and the
fact that deforming $\textrm{graph} u$ over a triangulation by increasing
$u_p$ at one point $p$ corresponds to a variation of $\textrm{graph} u$ with
variation vector field $\phi_p r_p$.

\subsection{Optimization techniques}
Now we have a problem of the following form:
\begin{eqnarray*}
  \label{eq:problem}
  \textrm{minimize}\quad &&\hat A(u)\quad u\in\mathcal{S}\approx \mathbf{R}^N \\
  \textrm{s.t.}\quad &&\hat V(u)-V_0 = 0
\end{eqnarray*}
with $\hat A (u)$ the area of $\textrm{graph} u$ over a fixed reference triangulation
and $\hat V (u)$ the discrete oriented volume between $\textrm{graph} u$ and the reference
triangulation as described above. 

Naturally one would consider a Lagrange method, that is computing the critical
points of the Lagrangian function
\[ L:\mathcal{S}\times\mathbf{R}\rightarrow\mathbf{R}:
(u,\lambda)\mapsto \hat A(u) - \lambda\left( \hat V(u) -V_0\right)\,. \]
However, since a local minimum of (\ref{eq:problem}) does not correspond to a
minimum of $L$ but merely to a critical point of $L$, minimization methods are
not applicable. Thus solving the critical point equation directly with a
Newton method is the only alternative in view, but this is not desirable,
since it would involve second derivatives of the metric $g$.

For numerical optimization however, there is a better method, namely the
augmented Lagrangian method. This considers the following penalized
Lagrangian function
\[ L_\rho(u,\lambda) = \hat A(u) - \lambda\left( \hat V(u) -V_0\right)
+\frac{\rho}{2}\left( \hat V(u) -V_0\right)^2 \]
with a \emph{penalty parameter} $\rho$. The following theorem can be proved
using standard calculus methods on the submanifold generated by the
constraint. However, a more elementary proof can be found in \cite[Theorem 12.2.1]{Fletcher:1981}.
\begin{theorem}
  \label{thm:auglag}
  If $\hat A$ and $\hat V$ are $C^2$, $u^*$ is a solution to
  (\ref{eq:problem}), $\lambda^*$ is the exact value of the Lagrange
  parameter, and the Hessian of $\hat A$ at $u^*$ is positive in directions
  perpendicular to $\textrm{grad} V$, then $u^*$ is a critical point of
  $L_\rho(\cdot,\lambda^*)$ and there exists $\rho_0$ such that for all
  $\rho>\rho_0$ the Hessian of $L_\rho(\cdot,\lambda^*)$ is positive definite,
  that is, $u^*$ is a strict local minimum of $L_\rho(\cdot,\lambda^*)$.
\end{theorem}
Note that in the analytic case, theorem~\ref{thm:hy} implies that the
conditions of this theorem hold. A CMC-surface is therefore a local
minimum for the analytic penalized Lagrangian for suitable penalty and
Lagrange parameters. For the discrete case we can still assert the regularity
assumption, but we unfortunately do not know about stability.

To solve (\ref{eq:problem}) we minimize the augmented Lagrangian. To find the
Lagrange parameter to the desired volume the following algorithm is used:
\begin{algo}
\item $\rho \leftarrow \textrm{some value}>\rho_0$
\item $\lambda_0\leftarrow\textrm{good initial guess}$
\item $k\leftarrow 0$
\item {\bf repeat} 
  \begin{algo}

  \item Minimize $L_\rho(\cdot,\lambda_k)$ to get approximative solution $u_k$
  \item $\lambda_{k+1} \leftarrow \lambda_{k} - \rho \left(\hat V(u_k) - V_0\right)$
  \item $ k \leftarrow k+1 $
  \end{algo}
\item {\bf until} $|\lambda - \lambda^*| < \eps $
\end{algo}
If the conditions of theorem~\ref{thm:auglag} hold, then this algorithm
converges locally in $(\lambda_k)$ to $\lambda^*$ and then the $(u_k)$
converge to $u^*$. Convergence improves for $\rho\rightarrow\infty$. For a
proof of this fact cf. \cite{Powell:1969}.

A ``good initial guess'' for $\lambda$ can be obtained from the Euclidean
situation. If a surface is considered that has a radius of approximately $r$ then
$2/r$, the Euclidean mean curvature of a sphere of radius $r$, is a good
choice, but in practice, starting with $\lambda=0$ also works well.

Each pair $(u_k,\lambda_k)$ obtained by one step of the above algorithm
corresponds to a discrete CMC-surface and its discrete mean curvature, of
course not with $V(u_k) = V_0$, but rather $V(u_k) = V(u_k)$. This explains
why not many steps have to be performed when one just wants to find
CMC-surfaces together with their mean curvature and area, but not necessarily a
particular enclosed volume.

The parameter $\rho$ should neither be chosen too small nor too big, since a
value too small gives that the minimization method diverges and a value too big
gives a bad condition of the minimization problem. An estimate for $\rho_0$
can be obtained by estimates for the Jacobi operator from Huisken-Yau
\cite{HY:1996}. These estimates indicate that $\rho_0$ decays like $r^{-3}$,
where $r$ is the radius of the surface considered.  But the condition of the
problem is not affected if $\rho$ is chosen to decay like $r^{-2}$ since that
is the scaling of the other terms of the Hessian matrix of $L_\rho$, which
controls the condition of the problem. Therefore the exact choice of
$\rho$ is not very critical, but decreasing $\rho$ should be attempted.

The penalized Lagrangian method is used to transform the constrained problem
into an unconstrained minimization. The method we chose to numerically solve
this is the conjugate gradient method. Although the conditions of
theorem~\ref{thm:auglag} imply local convergence of such a method, the rate of
convergence depends on the basis chosen for $\mathcal{S}$. The nodal basis is not a
very good choice, since the number of steps to make increases proportionally
to the number of points of the triangulation and would therefore give an
algorithm with quadratic time complexity.

The similarity of the problem to the solution of linear elliptic PDEs suggests
that one try methods from this field that have proven to give good convergence
rates. The choice made here is to use hierarchical bases that give the CG
method multigrid like speed while being very simple to implement. Using this
preconditioner, an overall time complexity of $\mathcal{O}(N\log N)$ for linear
problems is achieved with $N$ being the number of vertices of the
triangulation. Section~\ref{s:performance} describes some examples and shows
that the speedup is significant, and reduces complexity even in the nonlinear
case.

Optimization methods are presented in Fletcher
\cite{Fletcher:1980,Fletcher:1981}. For details on hierarchical bases see Bank,
Dupont and Yserentant \cite{BDY:1988} and Yserentant
\cite{Yserentant:1986a,Yserentant:1992}.

\subsection{Computation of single surfaces}
When a single surface has to be computed we can take the full advantages of
the hierarchical finite element representation, and use a cascading technique
for iteration as proposed by Bornemann and Deuflhard \cite{BD:1996}. We start
with a given coarse triangulation and iterate to get a first coarse
approximation to the surface. Then we refine this triangulation by dividing
each triangle into four new triangles. The surface obtained from the coarse
grid iteration then can be used as initial value to the fine grid iteration.
This procedure can be continued until the desired resolution is reached. This
\emph{cascading technique} gives an immense speedup.

\subsection{Computation of foliations}
When the method is used to compute whole families of constant mean curvature
surfaces close to Euclidean spheres it is possible to use the previously
computed surface as initial data for the next iteration.  To do that, one has
to produce a sequence of appropriate reference surfaces and volume conditions
that give a reasonable sequence of surfaces.

The following is suggested in a situation when one knows one discrete constant
mean curvature surface $\mathcal{T}$ that is centered. In this case, the
radial direction on this surface can be used as the variation direction of the
vertices.

To produce a new surface at distance $r$ further out (in), $\mathcal{T}$ should be
replaced by the graph of the constant function $r$ ($-r$) over $\mathcal{T}$. This
graph then will be the new reference surface $\tilde\mathcal{T}$. 

The constrained minimization problem to consider now is to minimize $\hat A$
while keeping $\hat V = 0$. This leads to an algorithm where not much volume
is inbetween the reference surface and the unknown constant mean curvature
surface, which is desirable since this increases the accuracy of the discrete
volume functional. 

The problem of finding the first constant mean curvature surface can sometimes
be solved by simply minimizing area without the constraint. This will give a
discrete minimal surface, which can serve as starting surface. However, not
all manifolds have a single spherical minimal surface that can be used.

\begin{figure}
  \centering
  \includegraphics[width=.4\linewidth]{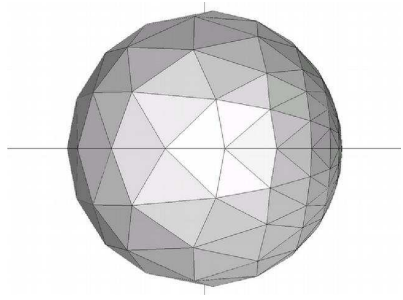}
  \caption{A translated triangulation has reduced resolution. This is the final iterate of
    an area minimization in Schwarzschild, started from a translated sphere.}
  \label{fig:distort}
\end{figure}

Another method to get a starting surface is to start with a Euclidean sphere
of big radius, use its normal direction as variation direction and solve the
constrained problem. However, if a significant translation occurs, the solver
should be restarted using a coordinate sphere around the computed center as
initial surface, since the distortion in the triangulation reduces the
resolution, for illustration see figure~\ref{fig:distort}. Alternatively one
could use adaptive mesh refinement here.

\subsection{Generalizing the Algorithm}
\label{s:generalized}
The algorithm described above is based on the variational structure of the
equation \[H = \textrm{const}\,.\]
In view of numerical relativity it would be very interesting to find surfaces
$\Sigma$ that solve the equations
\begin{equation}
  \label{eq:hpmp}
  H \pm P = \textrm{const} 
\end{equation}
with $P=\tr_\Sigma K = \tr_M K - K(\nu,\nu)$, $\nu$ the unit normal to
$\Sigma$, and $(M,g,K)$ are an initial data set for the Einstein equations.

However these equations are not the Euler-Lagrange equations for any
functional considering surfaces in $M$. This is due to the fact that the term
$P$ depends on the normal of the surface in consideration. In contrary the
equation
\[ H + H_0 = \textrm{const}\]
where $H_0 : M \to \mathbf{R}$ does not depend on the normal of the surface is the
Euler-Lagrange equation for a constrained minimization problem, in the sense
that it is satisfied on the boundary $\del\Omega$ of a set $\Omega\subset M$
minimizing
\begin{eqnarray*} 
  && J(\Omega) = |\del \Omega| + \int_\Omega H_0 \,\rmd\mu \qquad
  \Omega\subset M\\
  \textrm{subject to}&& \int_\Omega \,\rmd\mu = V_0 \equiv \textrm{const}\,.
\end{eqnarray*}
For a given initial surface $\Sigma_0$ with normal vector field $\nu_0$ we may
therefore fix a function $P_0 = \tr K - K(\nu_0,\nu_0)$ and extend it to a
neighborhood of the initial surface by the condition that it does not change
along the radial directions considered in~\ref{s:surfacemodel}. Then we solve
the equation $H \pm P_0 = \textrm{const}$ using the procedure described before. An
iteration of this strategy will produce a sequence of surfaces and normals
that might converge. If they converge, then the limit is a solution of the
discrete version of equation \ref{eq:hpmp}.

The author currently works on implementing, testing and examining this
approach, but came to the conclusion that this preliminary outline for
extending this algorithm might be interesting for applications in numerical
relativity.
%
\section{Numerical Examples}
\label{s:examples}
In this section we present three examples based on metrics of the
Brill-Lindquist-Type. This is a conformally flat metric on $\mathbf{R}^3$ of the form
\[ g_{ij}(x) 
= \left(1 + \sum_{k=1}^N \frac{m_k}{|x-x_k|}\right) \delta_{ij} \]
This metric represents a spacelike slice containing $N$ Schwarzschild-Type
singularities at the points $x_k$ of mass $m_k$. The metric and its
derivatives were evaluated analytically at every requested point.

\begin{figure}
  \centering
  \begin{minipage}[t]{.24\linewidth}
    \centering
    \includegraphics[width=.9\linewidth]{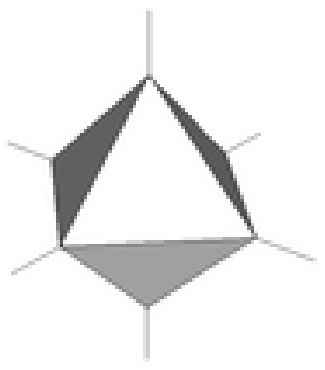}
  \end{minipage}
  \hfill
  \begin{minipage}[t]{.24\linewidth}
    \centering
    \includegraphics[width=.9\linewidth]{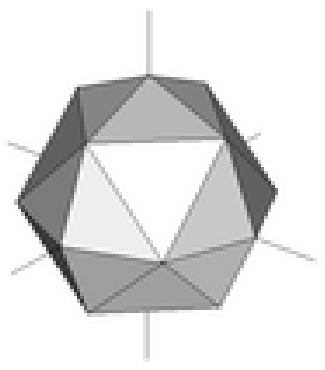}
  \end{minipage}
  \hfill
  \begin{minipage}[t]{.24\linewidth}
    \centering
    \includegraphics[width=.9\linewidth]{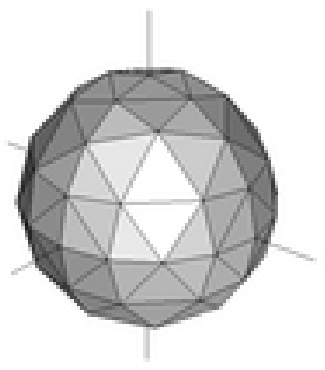}
  \end{minipage}
  \hfill
  \begin{minipage}[t]{.24\linewidth}
    \centering
    \includegraphics[width=.9\linewidth]{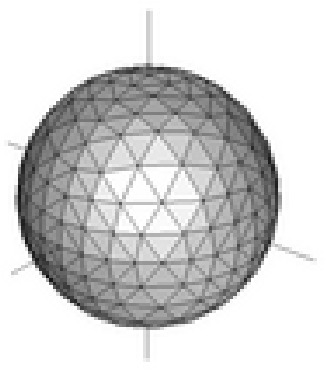}
  \end{minipage}
  \caption{Several refinement states of the octahedron.}
  \label{fig:octahedron}
\end{figure}

The triangulations used for the computations in this section are based on the
octahedron, that is the triangulation with vertices $\pm e_i$, $i=1,2,3$, and
regular refinement of it. To regularly refine a triangulation, every triangle
is divided into four subtriangles by introducing the midpoint of its edges as
new vertices to the triangulation. If these new points are projected to the
sphere one obtains the surfaces shown in figure~\ref{fig:octahedron}. Rescaled
and translated versions of these triangulations in different refinement states
will serve as starting surfaces.

All pictures of surfaces shown in this section were created using
\texttt{geomview} \cite{geomview}.

\subsection{Schwarzschild Solutions}
Here $N=1$, $m_1=1$ and $x_1=0$. This case was included to test the convergence
of the method. To compare the numerical results with the analytically known
results, introduce the following radius function for a surface $\Sigma$
\[ \tilde r = |\Sigma|^{-1}\int_\Sigma |x| d\mu_\Sigma \]
where $|\cdot|$ denotes Euclidean distance to the origin. Using the center of
gravity rule for numeric integration, this radius can be computed approximately,
which gives for each triangulation $\mathcal{T}$ a discrete radius $\hat r(\mathcal{T})$.

\begin{figure}
  \centering
  \begin{minipage}[t]{.48\linewidth}
    \includegraphics[width=\linewidth]{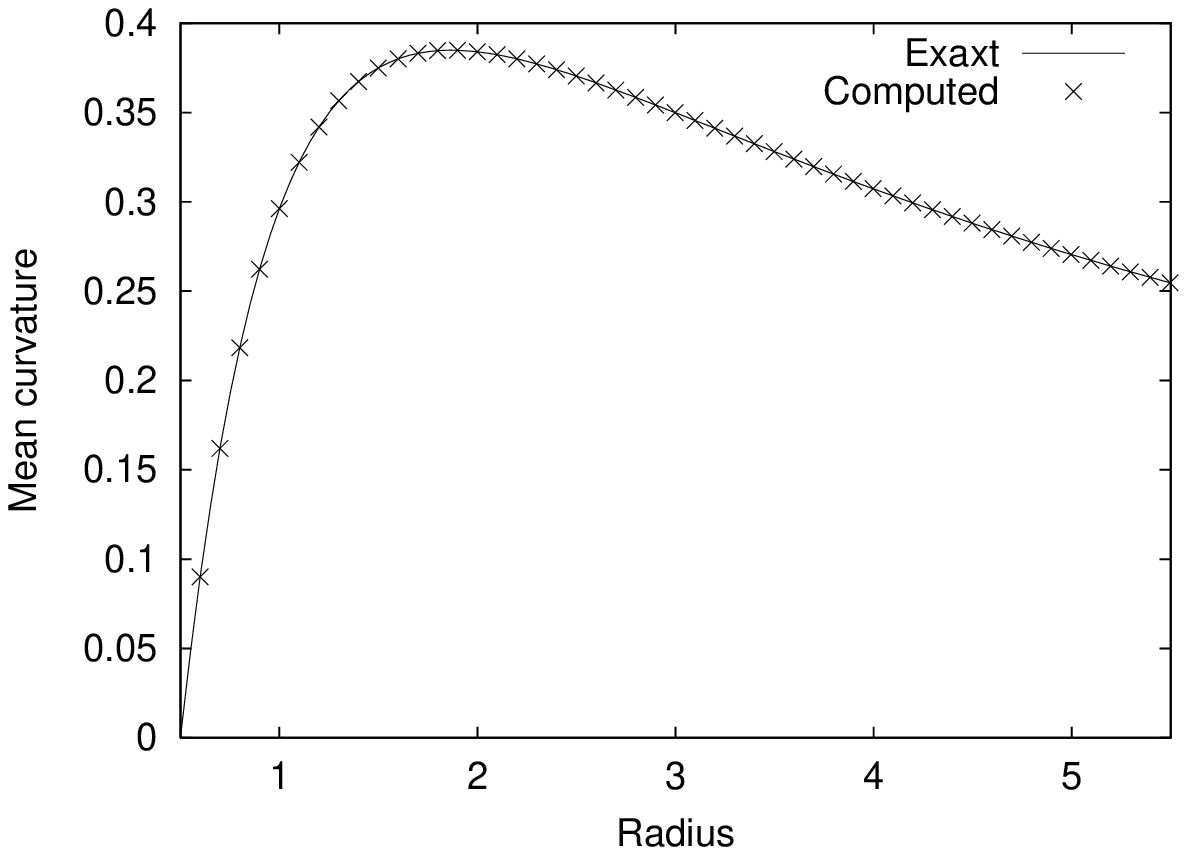}
    \caption{Exact and computed mean curvature compared for different radius.}
    \label{fig:schwarzschild-mc}
  \end{minipage}
  \hfill
  \begin{minipage}[t]{.48\linewidth}
    \includegraphics[width=\linewidth]{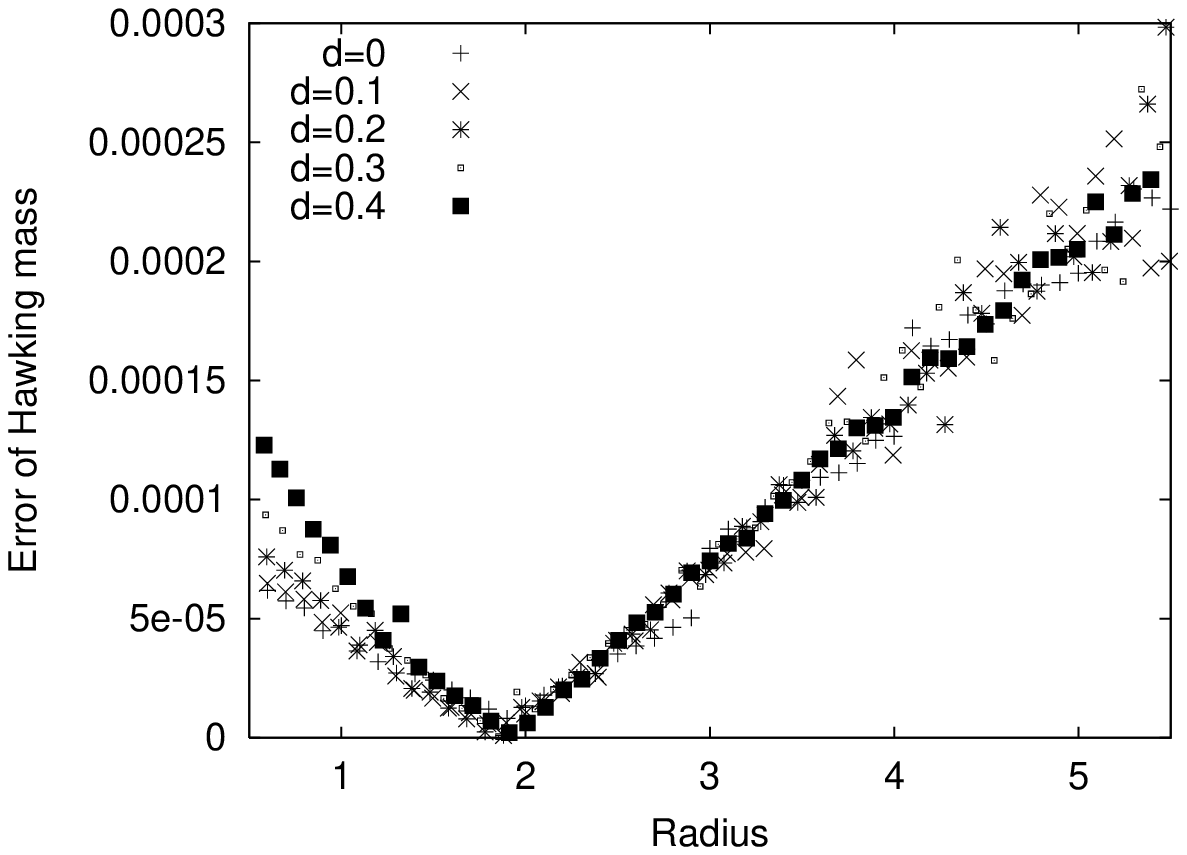}
    \caption{The difference of the Hawking mass for different translated
      initial spheres and the expected value 1.}
    \label{fig:schwarzschild-hawking}
  \end{minipage}
\end{figure}

Starting the method with the minimal surface and computing outward, one
obtains a family of triangulations $\mathcal{T}_k$, and a family of Lagrange
parameters $\lambda_k$ which can be interpreted as the constant mean curvature
of these triangulations. The exact mean curvature of a sphere of radius $r$ in
the spatial Schwarzschild metric is plotted as continuous line in
figure~\ref{fig:schwarzschild-mc} whereas the Lagrange parameters versus the
numerical radius of each triangulation is plotted as mark. This data results
from a resolution 6 triangulation.

Since a good approximation to the mean curvature is known, the Hawking mass
on constant mean curvature surfaces 
\[ m_H(\Sigma) = \frac{|\Sigma|^{1/2}}{(16\pi)^{3/2}}\left(16\pi - |\Sigma|H^2\right) \]
can easily be computed. Figure~\ref{fig:schwarzschild-hawking} shows the
absolute value of the difference of the Hawking mass of each surface and the
expected value 1. The initial surfaces were not centered here but translated
by $d=0,0.1,0.2,0.3,0.4$, resolution is again 6.

To test the ability of finding the center, the method was started with
Euclidean spheres of radius 20 and center $(d,0,0)$ for different values of
$d$. Then a CMC surface of equal volume was computed using the value 0.0005 as
penalty parameter. The Euclidean distance of the center of this surface to the
origin is shown in table~\ref{table:centerdev} for different resolutions and
different values of $d$.  The resolution number is the number of refinement
steps applied to each triangle. Resolution 3,4,5,6,7 correspond to
258,1026,4098,16386,65538 vertices.  This table shows nearly the expected
second order convergence, which corresponds to a factor of 4.
\begin{table}
\caption{Distance of the computed center to the origin for translated starting surfaces of radius 20 off
  center by distance $d$ and the associated convergence factor $f$ for $d=5$.}
\label{table:centerdev}
\begin{indented}
\item[]
\begin{tabular}{@{}cccccccc} 
\br
Resolution & $d=1$   & $d=2$   & $d=3$   & $d=4$   & $d=5$   & $f$ & $d=5$ \\ \mr
4          & 0.01791 & 0.03521 & 0.05394 & 0.07245 & 0.09161 & 4/5 & 3.93 \\
5          & 0.00455 & 0.00903 & 0.01382 & 0.01869 & 0.02329 & 5/6 & 3.95 \\
6          & 0.00112 & 0.00230 & 0.00343 & 0.00472 & 0.00589 & 6/7 & 3.98 \\
7          & 0.00030 & 0.00060 & 0.00086 & 0.00115 & 0.00148 & & \\ \br
\end{tabular}
\end{indented}
\end{table}

\begin{figure}
\centering
\begin{minipage}[c]{.48\linewidth}
\centering
\includegraphics[width=.8\linewidth]{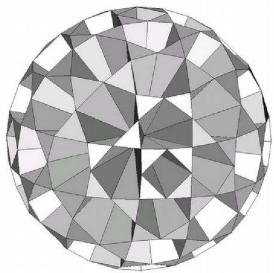}
\end{minipage}
\hfill
\begin{minipage}[c]{.48\linewidth}
\includegraphics[width=\linewidth]{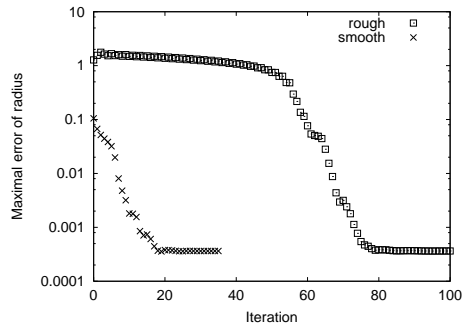}
\end{minipage}
\caption{A rough initial surface on the left and the the maximum of the
  absolute deviation of $r$ from 1/2 during an iteration for this surface. For
  comparison the data for an iteration with smooth data is shown.}
\label{fig:noise}
\end{figure}

A direct elliptic method fails to converge when started with a very rough
surface near the horizon, as is reported by Thornburg \cite{Thornburg:1996}.
To test this aspect of our method, we construct rough data by starting with a
sphere of radius 1/4 and refining up to level 5. The vertices that are created
by the last refinement step are moved outward to a radius of 3/4 such that the
starting surface oscillates around the minimal surface. In comparison to
spheres of radius 1/4 and 3/4 as starting surfaces, the iteration takes
significantly longer, about three times, but still converges. The diagnostic
quantity $E:=\max_\textrm{all vertices} |r-0.5|$ is plotted in
figure~\ref{fig:noise} for spheres of radius 1/4 and 3/4 and the rough surface
as initial surface. The iteration stopped, when the relative change in $E$
was less than 1/1000.

\subsection{Two singularities}
Here we examine a metric with $N=2$, $m_k=1$ and $x_k = \pm d e_1$. This space
contains two minimal surfaces, each one enclosing one of the singularities. 

The first test was to find the critical value $d^*$, such that the two
singularities are enclosed by common a minimal surface for $d<d^*$ but not for
$d>d^*$. Refinement up to level 7 indicates that $|d^* - 0.766362| \leq
0.000002$, which deviates by $0.02\%$ from the value $0.76619745$ reported by
Thornburg in \cite{Thornburg:2004}. The area of this minimal surface is
computed as $196.41579$, whereas Thornburg reported a value of $196.407951$
which deviates by $0.004\%$. We noticed however, that with increasing
resolution the value of $d^*$ and the area of the minimal surface decreased
with our procedure, therefore for larger resolution the values reported here
might become closer to the values of Thornburg.

\begin{figure}
\centering
\includegraphics[width=.6\linewidth]{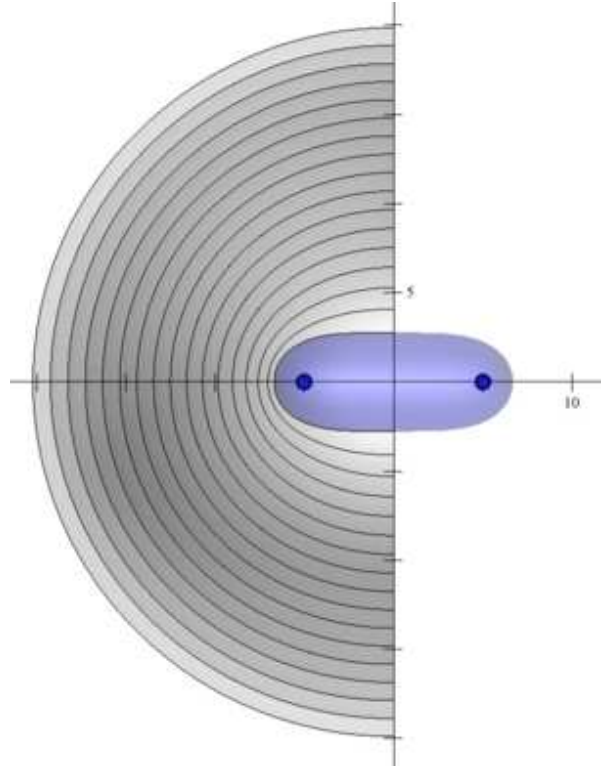}
\caption{The foliation for two singularities of mass 1 and distance 5 to the
  origin. The innermost surface is the last surface computed by the method,
  for smaller surfaces the algorithm seems to fail converging.}
\label{fig:th}
\end{figure}

For $d>d^*$, computation cannot start at a minimal surface, since this does
not exist. That is why for this problem the algorithm is started from a
Euclidean sphere with big radius centered at 0.  In each step the radius has
been reduced approximately by $1/10$. In figure~\ref{fig:th} every tenth
surface for $d=5$ is drawn. The two small surfaces are the minimal surfaces
enclosing the singularities.

\begin{figure}
  \centering
  \begin{minipage}[t]{.48\linewidth}
    \includegraphics[width=\linewidth]{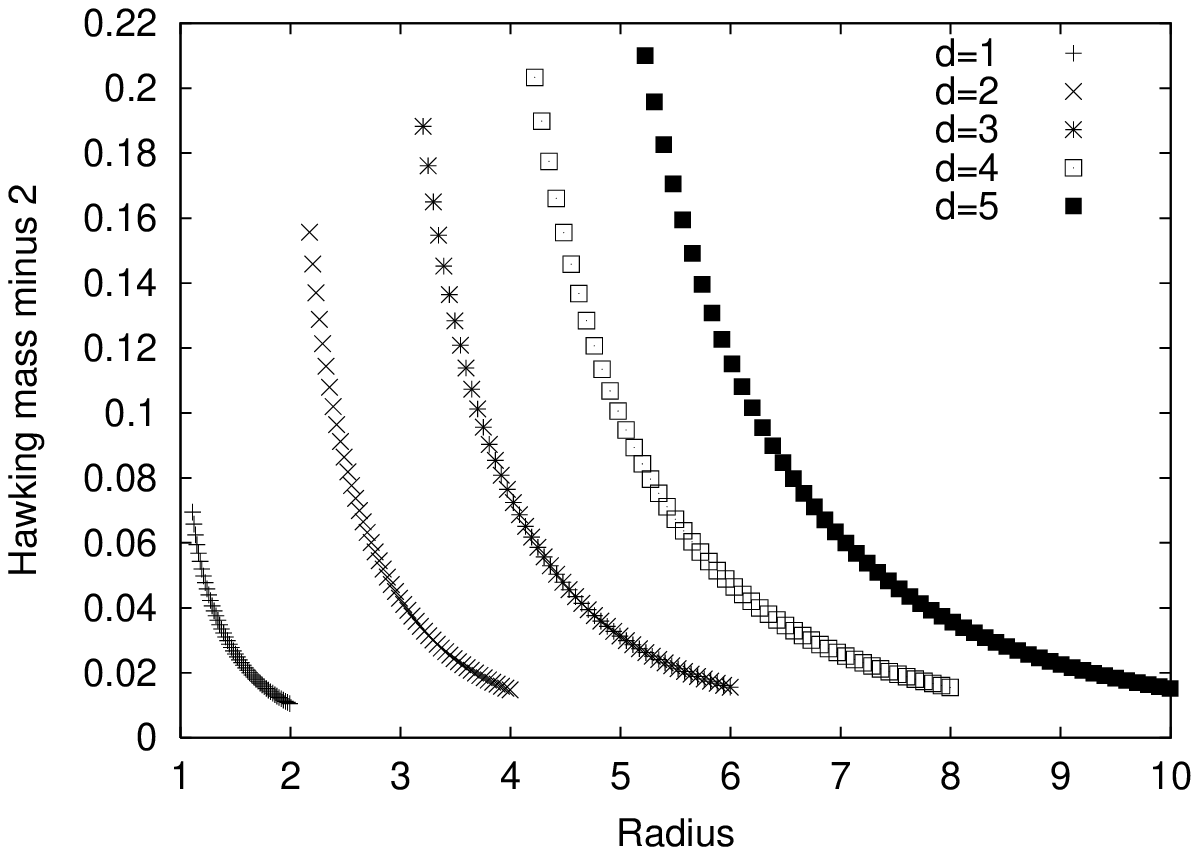}
    \caption{The error the Hawking mass for two singularities of mass~1.}
    \label{fig:th-hawking}
  \end{minipage}
  \hfill
  \begin{minipage}[t]{.48\linewidth}
    \includegraphics[width=\linewidth]{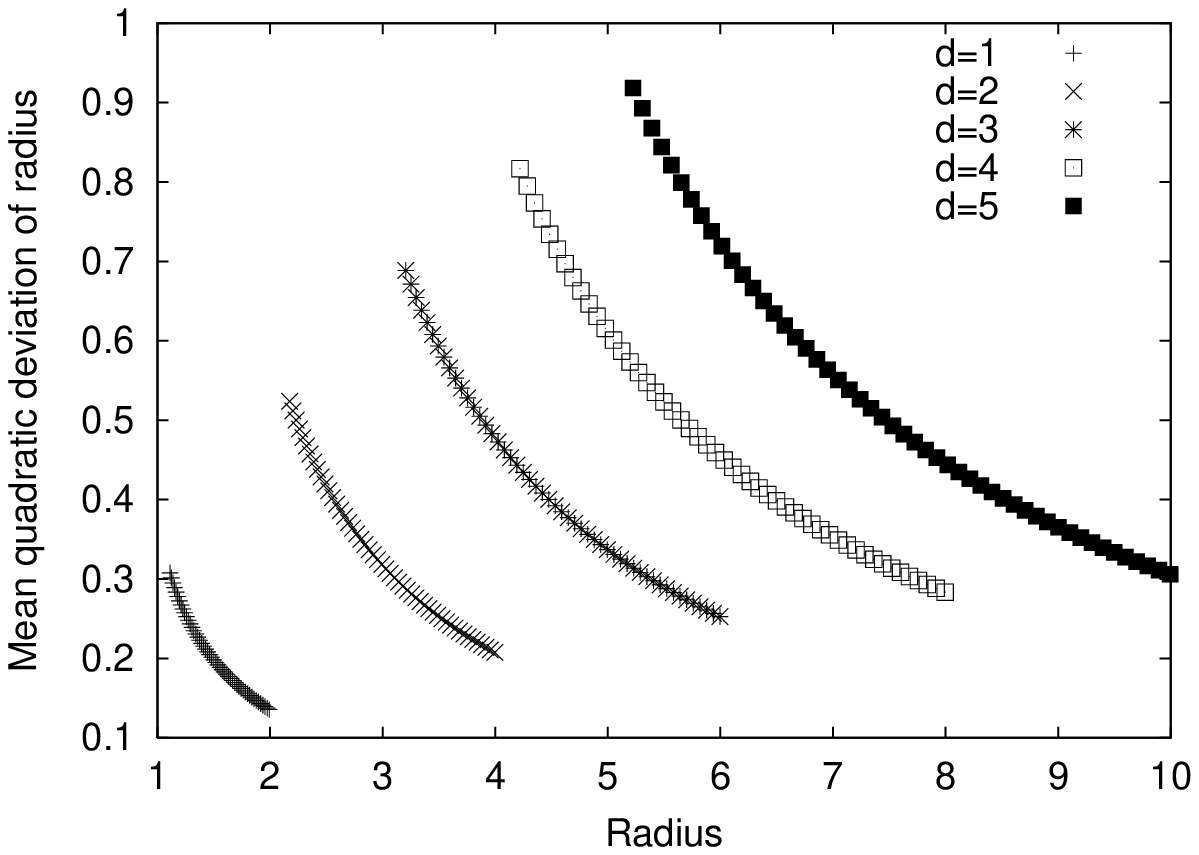}
    \caption{The mean quadratic deviation of the radius function
      for two singularities of mass~1.} 
    \label{fig:th-mqa} 
  \end{minipage}
\end{figure}

It is clear that for growing radius, the Hawking mass of the surfaces
approaches the ADM-mass, which in this case is the total mass of the two
singularities, namely 2. The difference of of the Hawking mass and the expected
value is plotted in figure~\ref{fig:th-hawking} for $d=1,2,3,4,5$.

The computed area, mean curvature and Hawking mass for $d=1$ of a surface with
approximate radius $5$ for different resolutions and the associated
convergence factor is shown in table~\ref{table:cf}.

Figure~\ref{fig:th-mqa} shows the mean quadratic deviation of the Euclidean
radius function from the expected radius. This figure shows that the surfaces
become rounder with growing radius as predicted by the theory.
\begin{table}
\caption{Area, mean curvature $H$, Hawking mass $m$ and associated convergence factors
  for the surfaces computed form a coordinate sphere with radius 5 in the
  $d=1$-metric for two singularities.}
\label{table:cf}
\begin{indented}
\item[]
\begin{tabular}{@{}cccccccc}
\br 
Resolution & Area & $H$ & $m$ & & $f_\textrm{Area}$ & $f_\textrm{H}$ &
$f_\textrm{m}$ \\ \mr
3 & 646.68121 & 0.1860585 & 1.9893651 & $\frac{3-4}{4-5}$ & 3.97 & 3.88 & 3.80 \\
4 & 649.42623 & 0.1857211 & 1.9926138 & \\
5 & 650.11737 & 0.1856342 & 1.9934694 & $\frac{4-5}{5-6}$ & 4.00 & 3.97 & 3.95 \\
6 & 650.29032 & 0.1856123 & 1.9936860 & \\ \mr
\end{tabular}
\end{indented}
\end{table}
%
\subsection{Three singularities}
Here we study a very symmetric Brill-Lindquist metric with three singularities
at $x_1 = 0$, $x_{2/3} = \pm e_1$ and masses $m_1 = 2$, $m_2=m_3 =2\sqrt{2}$.
This comes from the stereographic projection of a metric conformal to the
standard metric on $S^3\subset\mathbf{R}^4$. The conformal factor on the
sphere is the sum of the Greens functions to the conformal Laplacian at the
points $\pm(1,0,0,0)$ and $\pm(0,0,0,1)$, rescaled such that in the standard
stereographic projection of $S^3\setminus(0,0,0,1)$, the resulting metric has
the Brill-Lindquist form.

\begin{figure}
\begin{minipage}[c]{.48\linewidth}
\centering
\includegraphics[width=.8\linewidth]{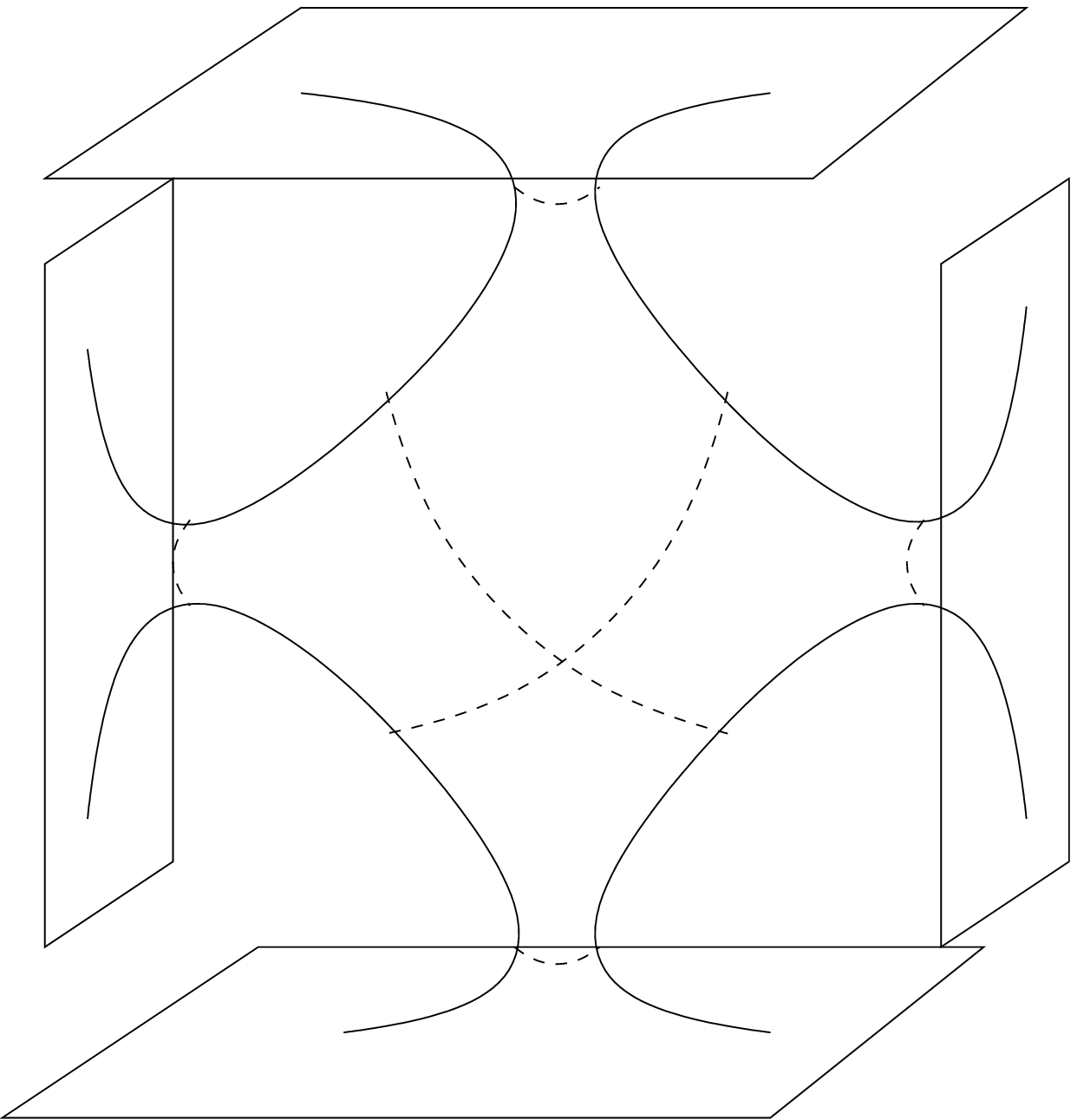}
\end{minipage}\hfill
\begin{minipage}[c]{.48\linewidth}
\includegraphics[width=\linewidth]{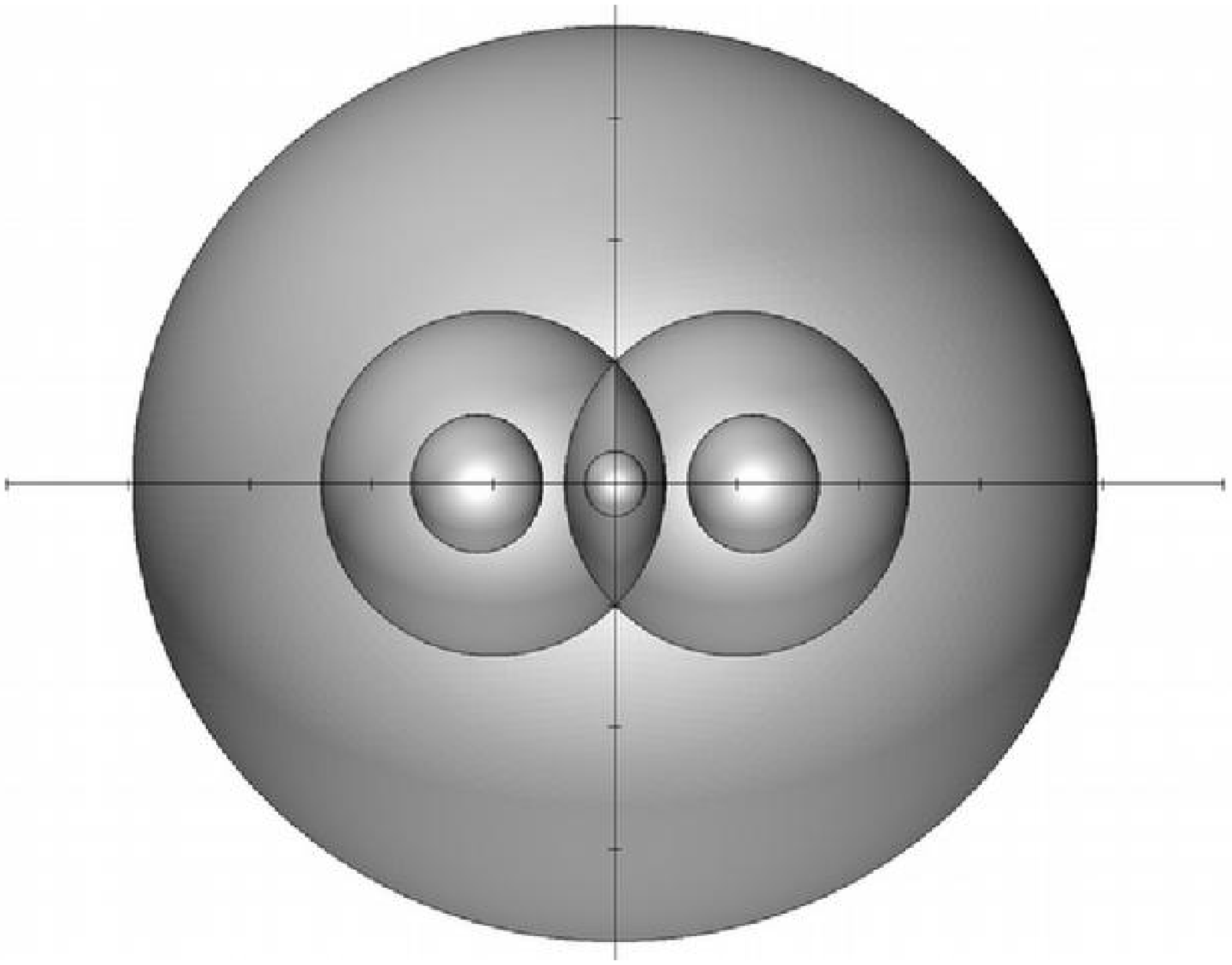}
\end{minipage}
\caption{\label{fig:conformal_s3} $S^3$ with four points blown to infinity by
  conformal rescaling, illustrated by $S^2$ on the left and in the stereographic
  projection with minimal surfaces on the right.}
\end{figure}

An analogous picture is shown in figure~\ref{fig:conformal_s3}, where it is
clarified that for each asymptotically flat end an outermost minimal surface
exists. Additionally there are two minimal surfaces resulting from the mirror
symmetry along the planes, that divide the picture in equal halves.  As it
turns out, these additional minimal surfaces are stable and can be found by
minimization.

The intersection of these minimal surfaces with the $x^1x^2$-plane is drawn
in figure~\ref{fig:conformal_s3} in the stereographic projection. The smaller
surfaces enclosing one singularity and the big surface enclosing all three
singularities correspond to the outermost minimal surfaces of the
asymptotically flat ends. The surfaces enclosing two singularities correspond
to the surfaces arising from the symmetry.
The area of the outermost minimal surfaces is given in
figure~\ref{table:conformal_area}. The surface named \emph{(R)ight} encloses the
singularity at $(1,0,0)$, the surface called \emph{(M)iddle} encloses the
singularity at the origin, and the surface called \emph{(O)uter} encloses all
three singularities. These areas should be the same due to symmetry. Therefore
we compute the convergence factors of the differences to zero and obtain, as
also shown in figure~\ref{table:conformal_area}, perfect second order
convergence.
\begin{table}
  \caption{The area of two different minimal surfaces and the convergence of its
    difference.}
  \label{table:conformal_area}
  \begin{indented}
  \item[]
  \begin{tabular}[t]{@{}ccccccc}\br
    Resolution & R    & M   & O     & f & R-M &
    R-O \\ \mr
    3          & 2966.95 & 2970.59 & 2970.84 & 3/4 & 4.10         & 4.09\\
    4          & 2951.64 & 2952.53 & 2952.59 & 4/5  & 4.02         & 4.02\\
    5          & 2947.79 & 2948.01 & 2948.03 & 5/6  & 4.00         & 4.00\\
    6          & 2946.83 & 2946.88 & 2946.89 & 6/7  & 4.00         & 4.00\\
    7          & 2946.59 & 2946.60 & 2946.60 \\\br
  \end{tabular}
  \end{indented}
\end{table} 

This symmetry enables us to compute these surfaces exactly. In $\mathbf{R}^4$
the surfaces are the intersections of $S^3$ with the planes $y^1=\pm y^4$ and
project to the surfaces given by the quadratic equation
\[ Q(x) := |x|^2 \pm 2x^1 - 1 = 0 \]

\begin{figure}
\centering
\includegraphics[width=.6\linewidth]{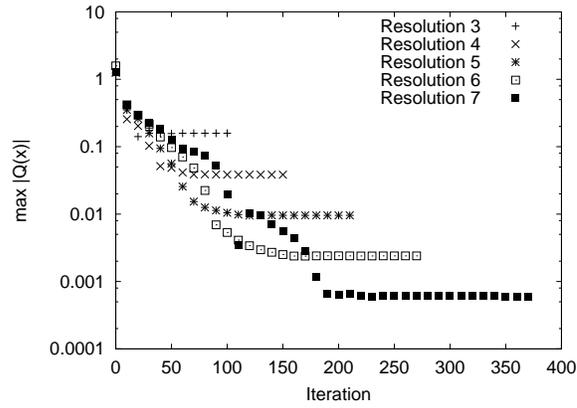}
\caption{The maximum value of $|Q(x)|$ for iterations in different resolutions.}
\label{fig:conformal_deviation}
\end{figure}

In figure~\ref{fig:conformal_deviation} we plot $\max_\textrm{all vertices}
|Q(x)|$ for an iteration started on a sphere with radius 1 centered at
$(0.5,0,0)$ for different resolutions. We again find quadratic
convergence for increasing resolution.

\subsection{Kerr Solutions}
\begin{figure}
  \begin{minipage}{.48\linewidth}
    \includegraphics[width=\linewidth]{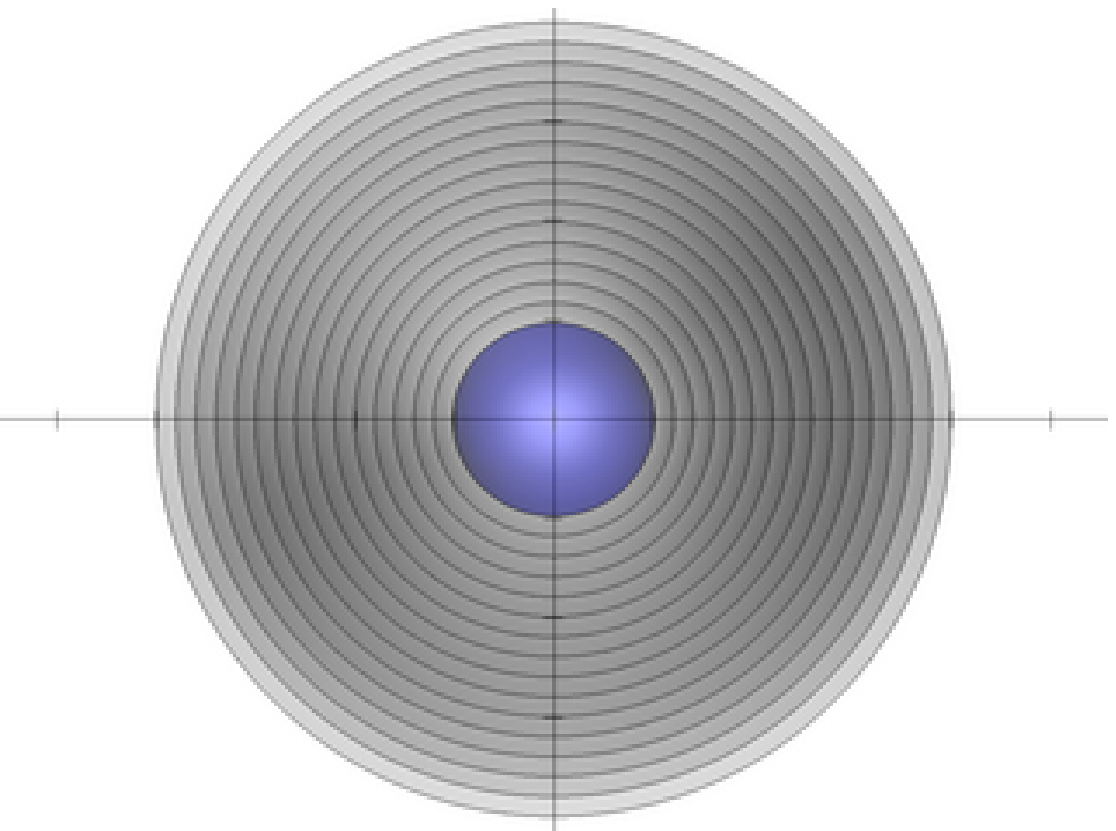}
  \end{minipage}
  \hfill
  \begin{minipage}{.48\linewidth}
    \includegraphics[width=\linewidth]{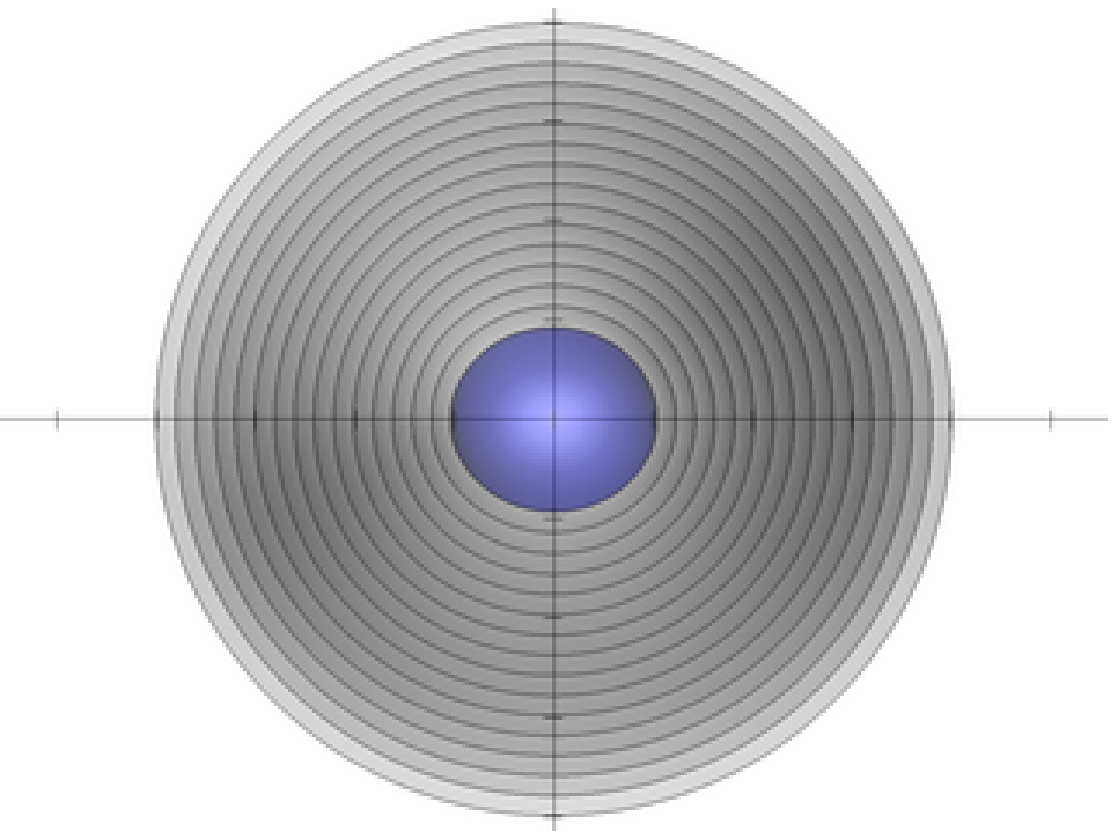}
  \end{minipage} \\[.5cm]
  \begin{minipage}{.48\linewidth}
    \includegraphics[width=\linewidth]{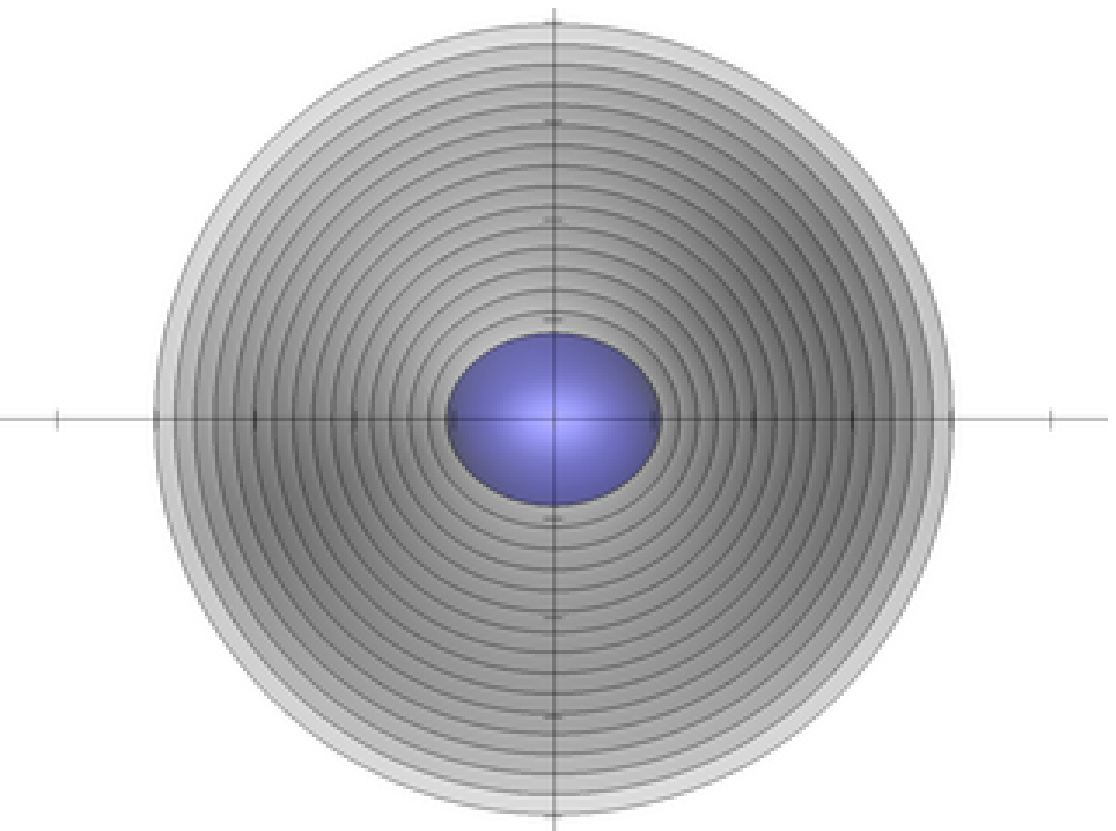}
  \end{minipage}
  \hfill
  \begin{minipage}{.48\linewidth}
    \includegraphics[width=\linewidth]{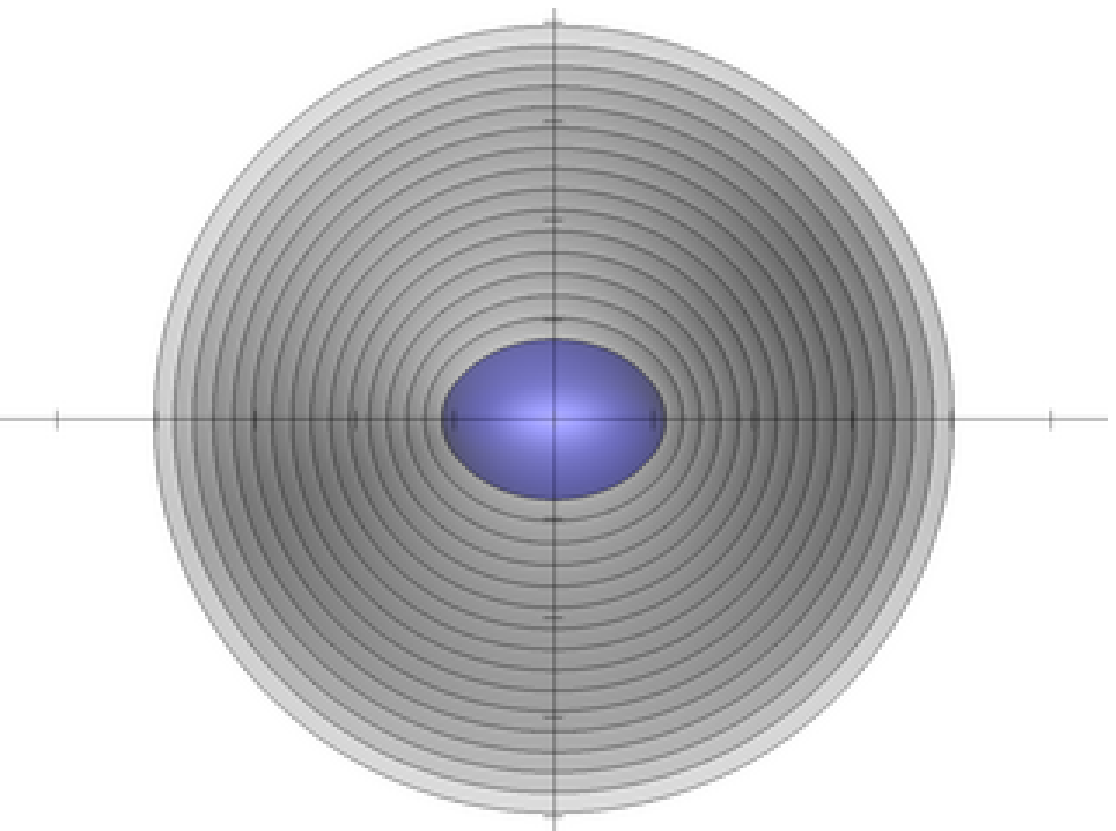}
  \end{minipage}
  \caption{The intersection of the constant mean curvature surfaces with the $x_1x_2$-plane
    in Kerr for $a=0.3$, $a=0.5$, $a=0.7$ and $a=0.9$. The ticks on the axes
    are in unit distance, the area radii of the innermost surfaces are 13.28,
    14.38, 15.75 and 17.11, further surfaces could not be computed.}
  \label{fig:kerr_surfaces}
\end{figure}  
For this test, a $\{t=\textrm{const}\}$-slice of the Kerr metric in
Kerr-Schild coordinates is used. The associated 3-metric is given by
\[ g_{ij} = \delta_{ij} + 2Hl_i l_j \]
with 
\[ H=\frac{MR}{R^2+a^2\cos^2\theta} \qquad l_i =
\left(\frac{Rx^1+ax^2}{R^2+a^2}, \frac{Rx^2-ax^1}{R^2+a^2},\frac{x^3}{R}\right) \]
where 
\[ R^2 = \frac{1}{2}(r^2-a^2)+\sqrt{\frac{1}{4}(r^2-a^2)^2+a^2(x^3)^2} 
\qquad \cos^2\theta=\frac{(x^3)^2}{R^2}\]
and $r^2 = |x|^2$ is the Euclidean distance. This metric has an apparent
horizon at $\{R=R_+\}$, $R_+=M+\sqrt{M^2-a^2}$ if $a<1$.  Unfortunately in this slicing
the horizon is not a CMC-surface in contrast to the Boyer-Lindquist
$\{t=\textrm{const}\}$ slices, and cannot be found with the method presented
here.

\begin{figure}
  \centering
  \includegraphics[width=.6\linewidth]{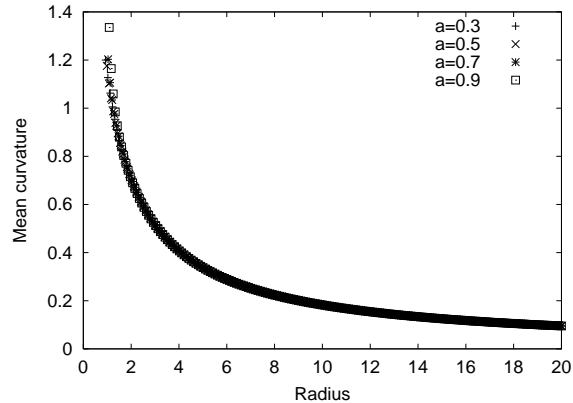}
  \caption{The mean curvature of the CMC-surfaces
    in Kerr for $a=0.3,0.5,0.7$ and $0.9$.}
  \label{fig:kerr_mc}
\end{figure}

\begin{figure}
  \begin{minipage}[t]{.48\linewidth}
    \includegraphics[width=\linewidth]{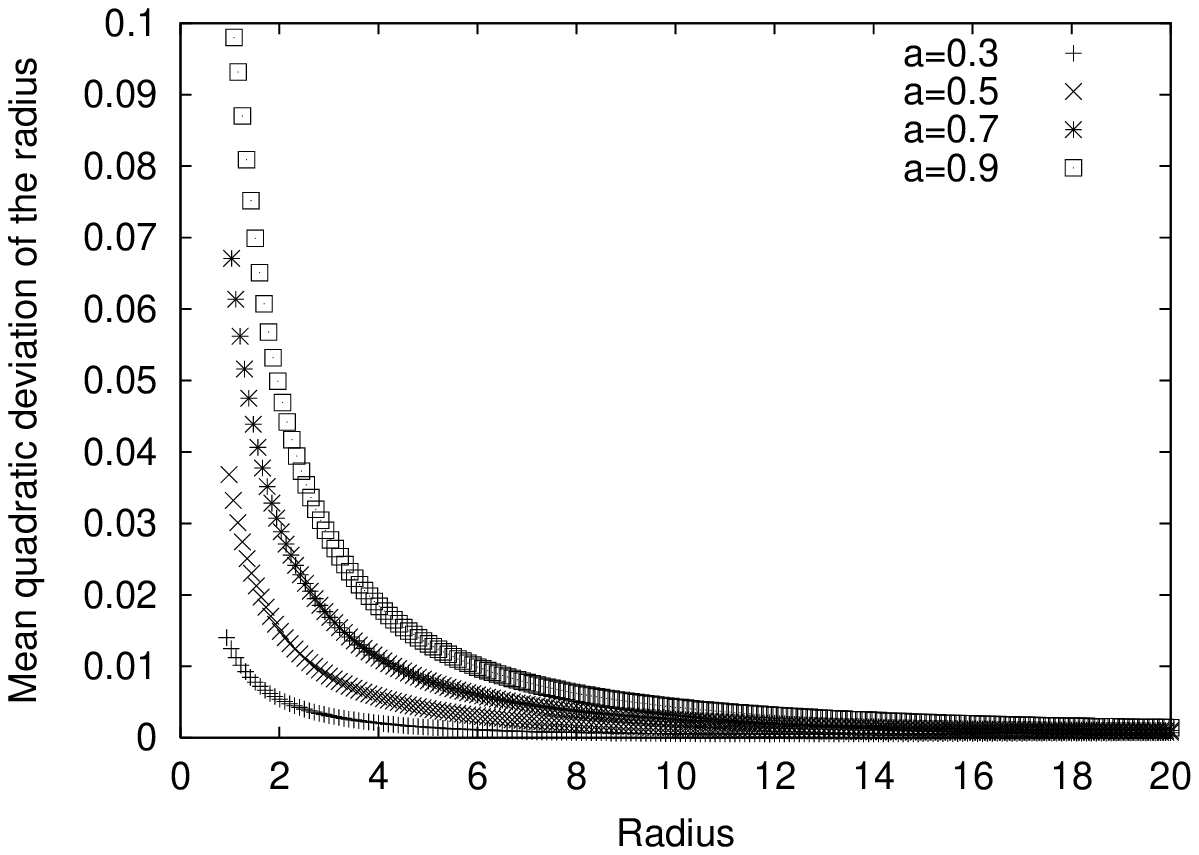}
  \end{minipage}
  \hfill
  \begin{minipage}[t]{.48\linewidth}
    \includegraphics[width=\linewidth]{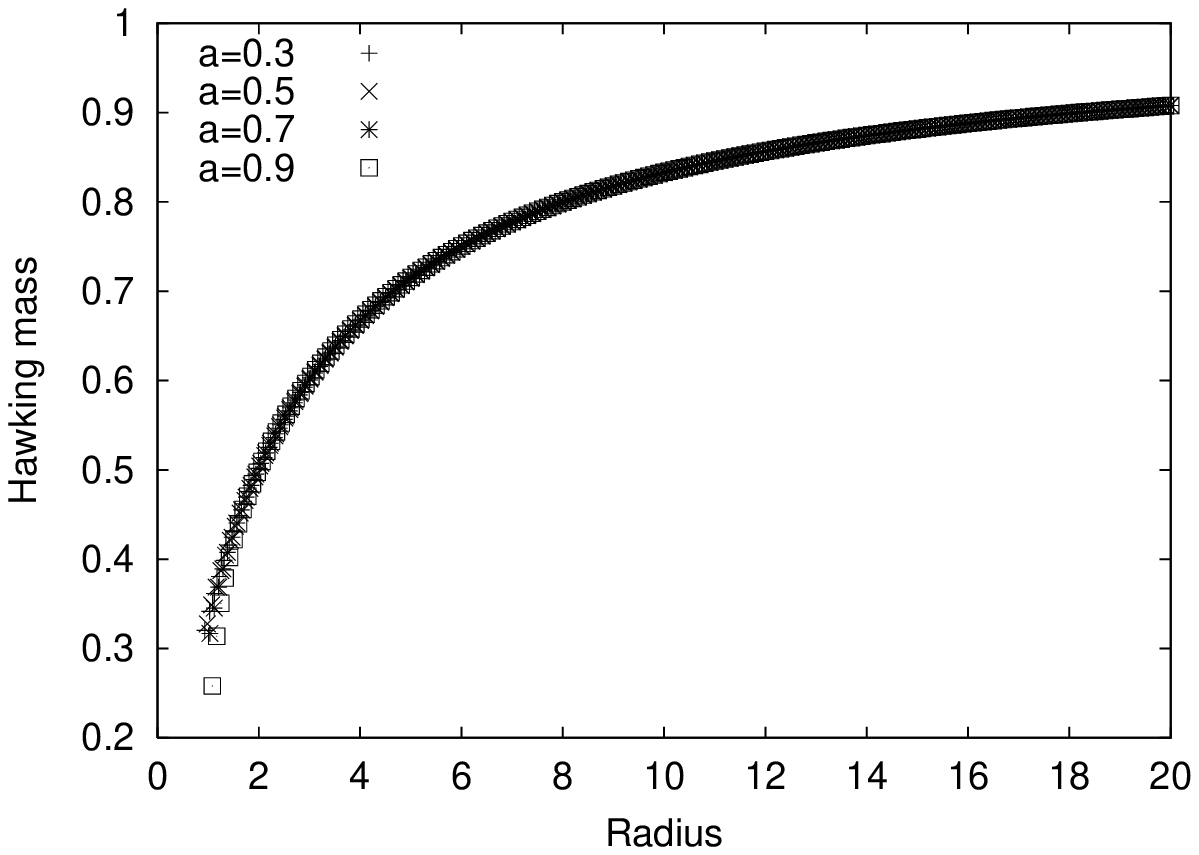}
  \end{minipage}
  \caption{Left, the mean quadratic deviation of the radius function of its
    mean value, right the hawking mass, of the CMC-surfaces in Kerr spacetime
    for $a=0.3,0.5,0.7$ and $0.9$.}
  \label{fig:kerr_mqa_hawking}
\end{figure}

We therefore start again at a large radius $r=20$ with a centered sphere and
compute inwards reducing the radius by 0.1 in each step. Some of the inner
surfaces for different values of $a$ are shown in
figure~\ref{fig:kerr_surfaces}. The last surface shown is the last surface
computed by the method. The mean curvature of some more surfaces is plotted
for these values of $a$ in figure~\ref{fig:kerr_mc}, while the mean quadratic
deviation of the radius function from constant radius and the hawking mass is
shown in figure~\ref{fig:kerr_mqa_hawking} for the same values of $a$. This
figure displays the remarkable fact that the hawking mass of the CMC-surfaces
is nearly independent of $a$, while not constant. The same shape of the graph
even holds for $a=0$, where the CMC-surfaces are perfectly round coordinate
spheres.  Figures~\ref{fig:kerr_mc} and~\ref{fig:kerr_mqa_hawking} show the
data of 190 surfaces that are nearly equidistant with distance 0.1, while in
figure~\ref{fig:kerr_surfaces} only 15 of the innermost surfaces with distance
0.2 are shown. The apparent horizon is shown in none of these pictures, since
it is not a CMC-surface. In the above plots we use the geometric area radius
$r_g=\sqrt{|\Sigma|/4\pi}$ on the horizontal axes.
\subsection{Performance}
\label{s:performance}
This section reviews the performance of the algorithm. All times reported here
have been measured on a personal computer with a single Athlon XP 2000+ CPU\@. The
first performance test is intended for comparing this method to others. The
metric used is the unit mass Schwarzschild metric $g_S =
(1+\frac{1}{2r})^4\delta$. The aim was to find the horizon, so area
minimization without volume constraint was attempted. The initial surface is
taken as discrete sphere of radius 1 with center $(0.3, 0, 0)$. We use a low
resolution of 3 to start and initially optimize. Then we refine to the
next resolution, take the surface obtained from the coarse grid iteration as
initial values for the fine grid optimization, and iterate this process until
the desired resolution is obtained.

\begin{figure}
  \centering
  \includegraphics[width=.6\linewidth]{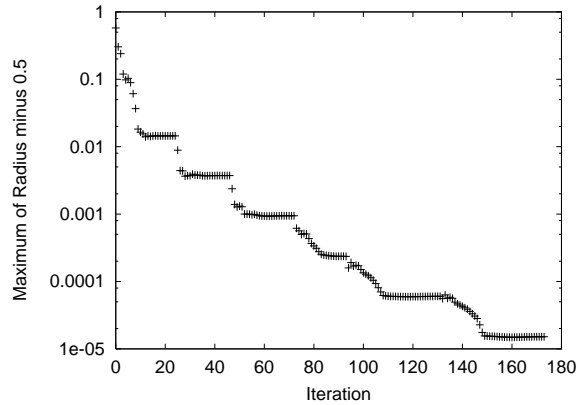}
  \caption{The maximum of the absolute value of the radius function minus 0.5 monitored during
    the iteration of a three to eight refinement steps run.}
  \label{fig:E}
\end{figure}

Since the horizon is a Euclidean sphere of radius $1/2$ we use $E:=
\max_\textrm{all vertices} |r-1/2|$ as diagnostic parameter. Iteration is stopped
when $E$'s relative change is less than $1/1000$ for more than four
iterations. Table~\ref{table:schwarzschild-error} shows the iterations elapsed
for each refinement level and the total time elapsed up to reaching the
result, including the coarse grid iterations. Note that the final values for
$E$ display nearly perfect second order convergence. Figure~\ref{fig:E} shows
the value of $E$ for each iteration of the algorithm with maximal refinement
level 8 starting with refinement level 3. The levels of the cascade
correspond to the convergent regime of the algorithm for the current
refinement level, while the edges occur when refinement has been done.
\begin{table}
  \caption{Elapsed Time and Iterations for the Schwarzschild test.}
  \label{table:schwarzschild-error}
  \begin{indented}
  \item[]
  \begin{tabular}{ccccc}\br
    Resolution & Number    & Iterations      & Total    & final value  \\
               & of points & on finest Level & Time [s] &   of  E     \\ \mr
    3          &       258 &              25 &     0.44 & 0.014472349 \\
    4          &      1025 &              22 &     3.22 & 0.003732145 \\
    5          &      4098 &              26 &       16 & 0.000945821 \\
    6          &     16386 &              21 &       58 & 0.000236132 \\ 
    7          &     65538 &              39 &      374 & 0.000060190 \\
    8          &    262146 &              42 &     1726 & 0.000015141 \\ \br
    \end{tabular}
  \end{indented}
\end{table}

\begin{figure}
  \begin{minipage}{.48\linewidth}
    \includegraphics[width=\linewidth]{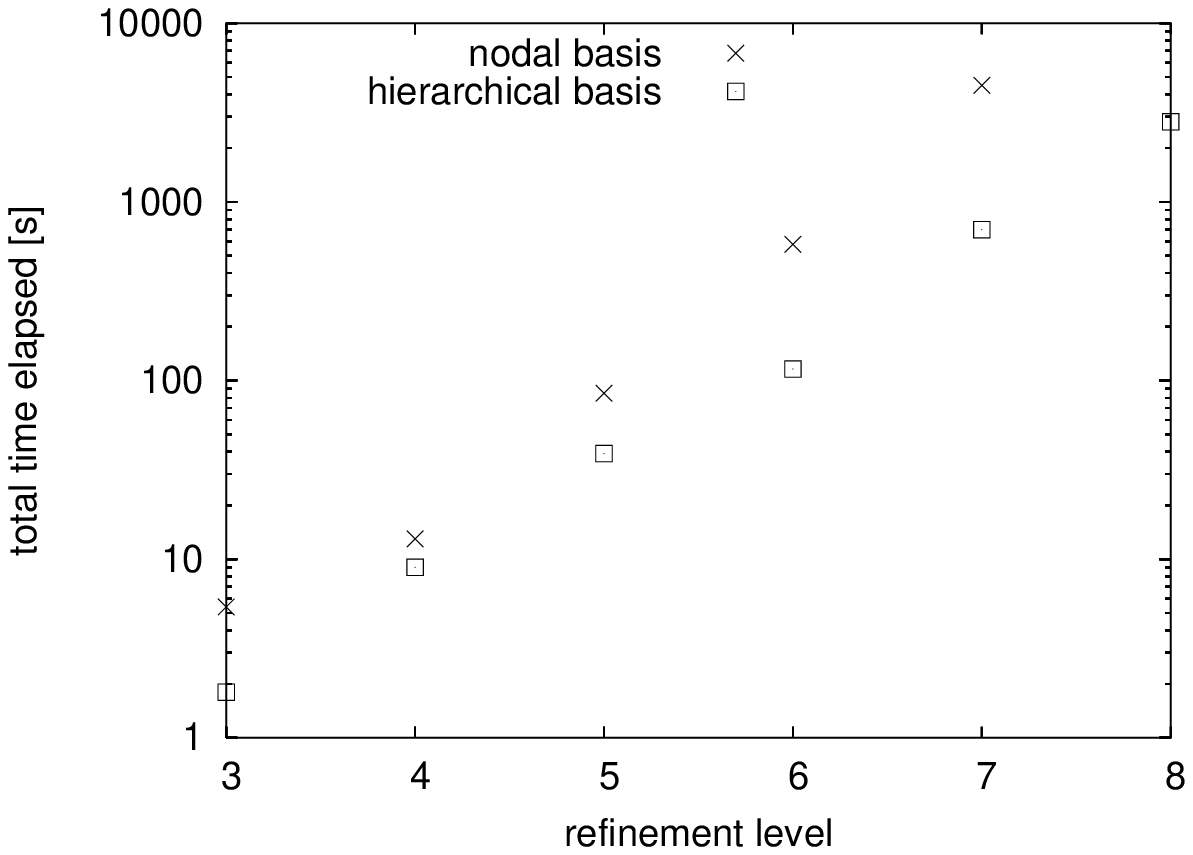}
  \end{minipage} 
  \hfill
  \begin{minipage}{.48\linewidth}
    \includegraphics[width=\linewidth]{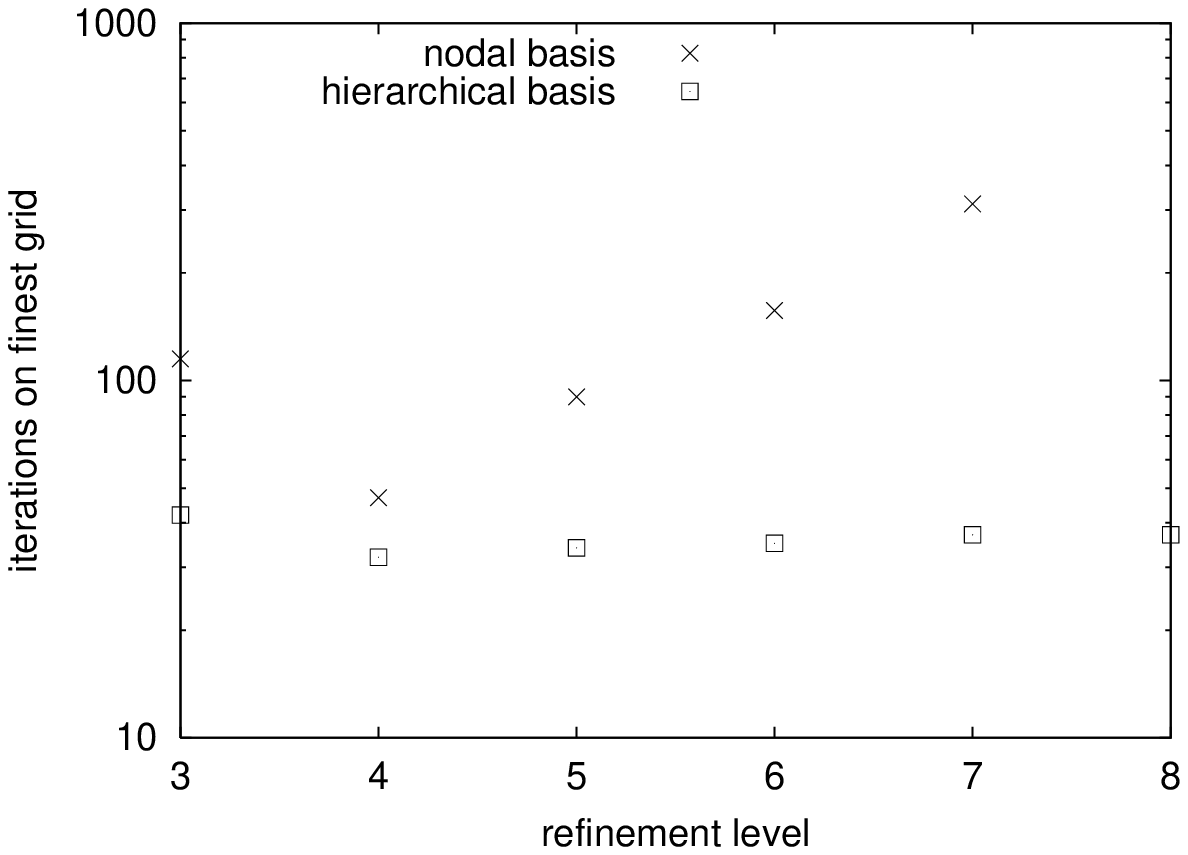}
  \end{minipage}
    \caption{Total time (left) and the number of iterations for the highest refinement
      level (right) plotted versus the number of refinement levels.}
    \label{fig:time_iter}
\end{figure}

To test the speedup of the hierarchical basis transformation, we go back to
Brill-Lindquist data with two singularities and $d=1$. We start with a large
coordinate sphere of radius $10$, translated by $1$ and test the time used to
center this sphere to a constant mean curvature surface using the nodal and
hierarchical basis. We again used the cascading technique with initial
refinement level 3. As diagnostic parameter we use the the $l^2$-gradient norm
times two to the power of the number of refinement levels, which is nearly
independent of the refinement level. The stopping condition was to reduce this
value by a factor of 2000. We chose to do only one minimization and therefore
started with the previously computed Lagrange parameter $\lambda=0.13531$. The
total time elapsed and the number of iterations in the highest resolution is
plotted in figure~\ref{fig:time_iter}. We see a substantial improvement in the number
of iterations involved and the elapsed time.

\subsection{Comparison to other methods}
Comparison of the speed of this method to other methods available is
very difficult at this stage. This is mainly due to the following
three issues. At first there are no standardized tests nor
standardized platforms on which to test nor agreements with what
accuracy to stop, which are general problems. Second, the figures
presented by Schnetter \cite{Schnetter:2003} and Thornburg
\cite{Thornburg:2004} do not refer to the computation of CMC-Surfaces
but to the computation of apparent horizons, which is a different
equation in general and especially in the cases considered by
them. And third, as stated before, our method uses an explicit
evaluation of the metric, whereas Schnetter and Thornburg use
interpolation of the metric from a given grid discretizing the metric
on the ambient three-manifold. As evaluation of the metric takes more
than half of the execution time in our case, it is hard to judge
whether we compare the underlying algorithms or just the different
methods of evaluating the metric.

However, in table~\ref{table:time-thornburg} we display some figures
taken from Thornburg to show in comparison to
table~\ref{table:schwarzschild-error} at least that the execution time
of our algorithm is in the same order of magnitude as the times given
by Thornburg. Thornburg tries to find the horizon in a boosted slice
of Kerr data with angular momentum $a=0.8$ and velocity $v=0.8$. The
numbers refer to execution time on a dual Intel Pentium IV 1.7 GHz which
executes his algorithm on one processor, where in contrast we worked
on a single AMD Athlon XP 2000+.
\begin{table}
  \caption{Values reported by Thornburg for horizon finding in a
  boosted Kerr time slice, with different resolutions.}
  \label{table:time-thornburg}
  \begin{indented}
  \item[]
    \begin{tabular}{cc}\br
    Number of points &  Total Time[s] \\ \mr
                 533 &  2.0 \\
                1121 &  4.2 \\
                2945 &  13  \\
                7905 &  43  \\ 
               25025 &  220 \\ \br
    \end{tabular}
  \end{indented}
\end{table}
\section{Conclusion}
We have presented an algorithm to compute constant mean curvature
surfaces based on a finite element discretization. The implementation
displays convergence, accuracy and speed rivaling the methods based on
finite differencing used in production runs. But still the
implementation by the author is academic in the sense that not much
effort has been spent on optimization and fitting the code into
standard environments such as \textsc{Cactus} for example.

However, creating a fully functional apparent horizon finder based on finite
element discretization, i.e. the extention of this algorithm to not
necessarily time symmetric initial data, is still a long way to go, but the
author's opinion is that the results presented here are very encouraging.

\ack 
I wish to thank Professor Huisken and Professor Yserentant for
getting me interested in this subject and for their constant
support. Additional thanks go to Erik Schnetter and Jonathan Thornburg
for interesting discussions on the subject. In addition I wish to
thank for financial support by the Sonderforschungsbereich 382 of the
DFG.

\section*{References}
\bibliographystyle{unsrt} 
\bibliography{references}

\begin{thebibliography}{10}

\bibitem{DKSS:2004}
Dreyer O, Krishnan B, Shoemaker D and Schnetter E
2003
{\it Phys. Rev. D}
{\bf 67}
024018

\bibitem{BS:2003}
Baumgarte T W and Shapiro S L
2003
{\it Phys. Rep.}
{\bf 376}
p~41--131

\bibitem{HY:1996}
Huisken G and Yau S T
1996
{\it Invent. Math.}
{\bf 124}
p~281--311

\bibitem{H:1998}
Huisken G
1998
{\it Trends in Mathematical Physics (Knoxville, TN, 1998)}
(Providence: American Mathematical Society)
p~299--306

\bibitem{KH:1995}
Krivan W and Herold H
1995
{\it Class. Quant. Grav.}
{\bf 12}
p~2297--2308

\bibitem{Bray:1997}
Bray H L
1997
{\it Ph.D. Dissertation}
Stanford University

\bibitem{Thornburg:1996}
Thornburg J
1996
{\it Phys. Rev. D}
{\bf 54}
p~4899--4918

\bibitem{Schnetter:2003}
Schnetter E
2003
{\it Class. Quant. Grav.}
{\bf 20}
p~4719--4737

\bibitem{Tod:1991}
Tod K P
1991
{\it Class. Quant. Grav.}
{\bf 8}
p~L115--L118

\bibitem{SHM:2000}
Shoemaker D, Huq M F and Matzner R A
2000
{\it Phys. Rev. D}
{\bf 62}
124005

\bibitem{BC:1984}
Barbosa J L and do~Carmo M 
1984 
{\it Math. Z.} 
{\bf 185} 
p~339--353

\bibitem{BC:1988}
Barbosa J L and do~Carmo M 
1988
{\it Math. Z.}
{\bf 197}
p~123--138

\bibitem{Leinen:1995}
Leinen P
1995
{\it Computing}
{\bf 55}
p~325--354

\bibitem{CL:1991} 
Ciarlet P G and Lions J-L, eds.  
1991 
{\it Handbook of numerical analysis. Vol II. Finite element methods. Part 1.}
(Amsterdam: North-Holland)

\bibitem{Fletcher:1981}
Fletcher R
1981
{\it Practical Methods of Optimization, Vol. 2}
(Chicester, New York, Brisbane, Toronto: John Wiley \& Sons)

\bibitem{Powell:1969}
Powell M J D
1969
in Fletcher R, ed. {\it Optimization} (Academic Press, London and New York)
p~283--298

\bibitem{Fletcher:1980}
Fletcher R
1980
{\it Practical Methods of Optimization, Vol. 1}
(Chicester, New York, Brisbane, Toronto: John Wiley \& Sons)

\bibitem{BDY:1988} 
Bank R E, Dupont T and Yserentant H 
1988 
{\it Numer. Math.}
{\bf 52} 
p~427--458

\bibitem{Yserentant:1986a}
Yserentant H
1986
{\it Numer. Math.}
{\bf 49}
p~379--412

\bibitem{Yserentant:1992}
Yserentant H
1992
{\em ICIAM (Washington 1991)} 
(Philadelphia: SIAM)
p~256--276

\bibitem{BD:1996}
Bornemann F A and Deuflhard P
1996
{\it Numer. Math.}
{\bf 75(2)}
p~135--152

\bibitem{geomview}
Phillips M et~al.
2000
{\it Geomview Manual}
(The Geometry Center -- http://www.geomview.org)

\bibitem{Thornburg:2004}
Thornburg J
2004
{\it Class. Quant. Grav.}
{\bf 21(2)}
p~743--766

\end{thebibliography}
\end{document}